\documentclass[12pt]{article}
\usepackage{amsmath,amssymb}
\usepackage{cite}

\newcommand{\al}{\alpha}
\newcommand{\ep}{\epsilon}

\newcommand{\la}{\lambda}
\newcommand{\La}{\Lambda}

\newcommand{\lb}{\lbrack}
\newcommand{\rb}{\rbrack}

\newcommand{\msc}[1]{\mbox{\scriptsize #1}}
\newcommand{\dsp}{\displaystyle}

\newcommand{\bc}{\mathbb{C}}
\newcommand{\br}{\mathbb{R}}
\newcommand{\bz}{\mathbb{Z}}

\newcommand{\dal}{\dot{\alpha}}

\newcommand{\cG}{{\cal G}}

\newcommand{\cO}{{\cal O}}
\newcommand{\cN}{{\cal N}}
\newcommand{\cM}{{\cal M}}

\newcommand{\CP}{\mathbb{C}{\bf P}}

\newcommand{\id}{\mbox{id}}

\newcommand{\Th}[2]{\Theta_{#1,#2}}
\renewcommand{\th}{{\theta}}
\newcommand{\tTh}[2]{\widetilde{\Theta}_{#1,#2}}

\newcommand{\Ch}[1]{\mbox{Ch}^{#1}}

\newcommand{\sNS}{\msc{NS}}
\newcommand{\stNS}{\widetilde{\msc{NS}}}
\newcommand{\sR}{\msc{R}}
\newcommand{\stR}{\widetilde{\msc{R}}}

\newcommand{\lieg}{{\bf g}}
\newcommand{\lieh}{{\bf h}}
\newcommand{\liem}{{\bf m}}

\newcommand{\fiber}[1]{\stackrel{#1}{\longrightarrow}}

\newcommand{\eqn}[1]{(\ref{#1})}
\newcommand{\Zb}{\mathbb{Z}}

\newcommand{\SO}{{\ensuremath{SO}}}
\newcommand{\SU}{{\ensuremath{SU}}}
\newcommand{\U}{{\ensuremath{U}}}

\newcommand{\hc}{g^{*}}
\newcommand{\hs}{h^{*}}
\newcommand{\Ocal}{{\cal O}}

\newcommand{\Mcal}{{\cal M}}
\newcommand{\nn}{\nonumber}
\newcommand{\gauss}[1]{\left[#1\right]}
\newcommand{\tri}{{\rm tri}}
\newcommand{\isi}{{\rm Isi}}
\newcommand{\Spin}{{\itshape Spin}}

\numberwithin{equation}{section}

\begin{document}

\vskip 7mm

\begin{titlepage}

\begin{flushright}
 \begin{tabular}{l}
  {\tt hep-th/0301164}\\
  UT-03-02\\
 \end{tabular}
\end{flushright}

 \vfill
 \baselineskip=20pt
 \begin{center}
 \centerline{\Huge  Supercoset CFT's for String Theories} 
 \vskip 5mm 
 \centerline{\Huge  on Non-compact Special Holonomy Manifolds}

 \vskip 2.0 truecm
\noindent{\it \large Tohru Eguchi, Yuji Sugawara and Satoshi Yamaguchi} \\
{\sf eguchi@hep-th.phys.s.u-tokyo.ac.jp~,~
sugawara@hep-th.phys.s.u-tokyo.ac.jp~, \\
yamaguch@hep-th.phys.s.u-tokyo.ac.jp}
\bigskip

 \vskip .6 truecm
 {\baselineskip=15pt
 {\it Department of Physics,  Faculty of Science, \\
  University of Tokyo \\
  7-3-1 Hongo, Bunkyo-ku, Tokyo 113-0033, Japan}
 }
 \vskip .4 truecm

 \end{center}

 \vfill{}
 \vskip 0.5 truecm

\begin{center}
 {\bf Abstract}
\end{center}
We study aspects of superstring vacua of non-compact special holonomy 
manifolds with conical singularities constructed systematically 
using soluble $\cN=1$ superconformal field theories (SCFT's). 
It is known that Einstein homogeneous spaces $G/H$ generate
Ricci flat manifolds with special holonomies on their cones 
$\simeq \br_+ \times G/H$,  when they are endowed with 
appropriate geometrical structures, namely, 
the Sasaki-Einstein, tri-Sasakian, nearly K\"{a}hler, and weak $G_2$ 
structures for $SU(n)$, $Sp(n)$, $G_2$, and $Spin(7)$ holonomies, 
respectively.  Motivated by this fact, we consider the string vacua 
of the type: $\br^{d-1,1} \times (\cN=1 ~ \mbox{Liouville}) \times (\cN=1 ~ 
\mbox{supercoset CFT on} \,\,G/H )$  where 
we use the affine Lie algebras of $G$ and $H$ 
in order to capture the geometry associated to an Einstein homogeneous space
$G/H$.
Remarkably, we find the same number of spacetime and worldsheet SUSY's 
in our ``CFT cone'' construction as expected from  
the analysis of geometrical cones over $G/H$ in many examples. 
We also present an analysis on the possible Liouville potential terms 
(cosmological constant type operators) 
which provide the marginal deformations resolving the conical 
singularities.

\vfill

\end{titlepage}


\section{Introduction}

String theories/M-theory on special holonomy manifolds 
developing conical singularities are of great importance
by several reasons: Firstly, they exhibit the non-perturbative 
quantum effects, such as gauge symmetry enhancements,
due to the appearance of light solitonic states
\cite{Strominger,Aspinwall,Witten}. 
Secondly, they are expected 
to provide frameworks to discuss the holographic dualities 
with the local or non-local interacting theories  defined on 
the asymptotic boundaries \cite{AMV,AW}, which naturally generalize the  
$AdS/CFT$-correspondence \cite{AdSCFT}.

It is known that an arbitrary $(m-1)$-dimensional 
Einstein space $X_{m-1}$ possesses a Ricci flat metric on 
its $k$-dimensional cone $C(X_{m-1})$ of the form 
\begin{eqnarray}
ds^2 = dr^2 + r^2 ds^2_{X_{m-1}} ~,
\label{cone metric}
\end{eqnarray}
where $r$ is the radial coordinate, and 
the special holonomies on $C(X_{m-1})$  
originate from the ``weak special holonomies''
on $X_{m-1}$ \cite{KW,AFHS,Gubser,MP}.  
To be more precise, the $SU(n)$, $Sp(n)$, $G_2$
and $Spin(7)$ holonomies on the cone $C(X_{m-1})$ are 
in one to one correspondence with the {\em Sasaki-Einstein\/} 
($m=2n$) \cite{Sasaki}, {\em tri-Sasakian\/} ($m=4n$) \cite{BG,GS}, 
{\em nearly K\"{a}hler\/} ($m=7$) \cite{Gray}
and {\em weak\/} $G_2$ ($m=8$) structures \cite{FKMS} on $X_{m-1}$, respectively
as proved in \cite{FK}\footnote
    {A good review for physicists on these mathematical concepts 
     is found in the paper \cite{AFHS}.} (See the table 1.).
This fact is very useful to systematically construct
special holonomy manifolds with conical singularities, 
because the Einstein homogeneous spaces $X_{m-1}=G/H$
endowed with these geometrical structures 
are well understood since the old days of Kaluza-Klein supergravity
(SUGRA) \cite{Witten-G,DFN,CR,PP,CDF,CRW,Romans} (and \cite{DNP} for a review) 
as well as from the mathematical literature mentioned above
\cite{Sasaki,BG,GS,Gray,FKMS,FK} and \cite{Aloff}.

On the other hand, there also exists extensive literature on 
the worldsheet approaches to these conical 
backgrounds in string theory.
Early literature for the conifold and $K3$-singularity
is \cite{GV} and \cite{OV}. 
More recent studies are given in \cite{ABKS,GKP,GK,Pelc} 
for the $SU(n)$-holonomies, emphasizing the role of 
$\cN=2$ Liouville theory \cite{KutS} and the holographically dual
descriptions based on the (wrapped) NS5-brane geometry (see also
\cite{HK}). All of these cases possess the worldsheet $\cN=2$
superconformal symmetry and have been discussed in 
\cite{ES1,Mizoguchi,Yamaguchi,NN} from the viewpoints 
of the ``non-compact extensions'' of Gepner models \cite{Gepner}. 
There are also several related results \cite{GVW,ShV,GGW,EWY}
from the spacetime view points, but with the RR-flux at infinity,
and also a work based on the Hybrid formalism \cite{hybrid}
is given in \cite{IK}. 

While these constructions in the case of the $\cN=2$ supersymmetry have been rather successful,
it is difficult to construct string vacua
on the conical backgrounds with 
$Spin(7)$ and $G_2$ holonomies, which possess at most the $\cN=1$
worldsheet SUSY. Partial attempts to construct 
the string vacua of $Spin(7)$ and $G_2$ holonomies with conical 
singularities have been given in \cite{ES3,SY2}. 
General structure of string theory on manifolds with $Spin(7)$ and $G_2$
was discussed in \cite{SV,F-O} in particular from the point of view of 
the existence of extended chiral algebras.
Structure of these chiral algebras has also been discussed 
in \cite{GN,Noyvert,SY1}.
There are also several results \cite{JoyceCFT}
for the CFT constructions of 
compact $G_2$ and $Spin(7)$ manifolds based on the geometrical 
method of Joyce \cite{Joyce}.

The main purpose of this paper is to give a systematic way 
of constructing special holonomy manifolds with conical 
singularities based on the solvable $\cN=1$ SCFT's, 
which may be regarded as a natural generalization of 
the construction in the  $\cN=2$ category mentioned above. 
Our strategy is quite simple:
{\em We formally replace an Einstein homogeneous space $X=G/H$
by an $\cN=1$ supercoset CFT $\cM=\left(G\times SO(n)\right)/H, 
\,(n=\dim G-\dim H)$ 
based on the affine Lie algebra of $G$ and $H$. 
$SO(n)$ stands for $n$ free fermions. 
We then add the $\cN=1$ Liouville 
sector in place of the radial degrees of freedom. } We may call 
our construction as the ``CFT cone" as opposed to the original 
geometrical cone construction.
Of course, one should keep in mind that the coset CFT (WZW model) of $G/H$
is not identical to the non-linear $\sigma$-model
with the target manifold $G/H$ because of the presence of NS $B$ field
in WZW models. 
Nevertheless, as we will see in the following, 
taking the supercoset CFT associated with the Einstein homogeneous 
space 
provides a very good anzats for the superstring vacua of special holonomy. 
We completely classify these coset constructions
(at least for the cosets $G/H$ with compact simple groups $G$), 
which include the models found in \cite{ES3,SY2}
as well as many of the vacua in the $\cN=2$ category presented in 
\cite{GV,OV,ABKS,GKP,GK,Pelc}.
Among other things, we will find that our CFT cone
approach leads to the right amount of worldsheet and spacetime 
SUSY's as expected from geometrical grounds in many examples.

~

This paper is organized as follows:
In section 2 we clarify the properties of conformal blocks 
characterizing the special holonomies in our CFT cone anzats. 
We also demonstrate the explicit constructions of spacetime supercharges.
In section 3, which is the main part of this paper,
we exhibit the classification of our string vacua with diverse spacetime
dimensions, and observe how we obtain the special holonomies which 
coincides (or differs) with those of the 
geometrical constructions of Ricci flat cones.
In section 4 we analyse the spectrum of possible cosmological constant type 
operators, which will provide the marginal deformations
resolving the conical singularities in backgrounds.
Section 5 is devoted to discussions on open problems.

~


\begin{table}[h]
 \begin{tabular}{|c|c|c|c|c|c|}\hline
 dim & name & holonomy & Killing spinor & $\sharp$ spacetime SUSY
 & base of the cone \\\hline\hline
 4 & hyper K\"ahler & $SU(2)$ & (2,0) & 16 & tri-Sasakian\\ \hline
 6 &Calabi-Yau & $SU(3)$ & (1,1) & 8 &  Sasaki-Einstein \\ \hline
 7 & $G_2$ & $G_2$ & 1 & 4 & nearly K\"ahler \\ \hline
 8 & hyper K\"ahler & $Sp(2)$ & (3,0) & 6 & tri-Sasakian \\ \hline
 8 &Calabi-Yau & $SU(4)$ & (2,0) & 4 & Sasaki-Einstein\\ \hline
 8 & $Spin(7)$ & $Spin(7)$ & (1,0) & 2 & weak $G_2$\\ \hline
 \end{tabular}
\caption{We summarize the relation between the special holonomies on 
 the cone $C(X_{m-1})$ and the geometrical structures on 
 the Einstein spaces $X_{m-1}$.}
\end{table}


\section{Superstring Vacua of Non-compact Special Holonomy Manifolds 
as the `Cone over SCFT's'}

\subsection{General Set Up and  Aspects of Spacetime SUSY}

In this paper we shall search for 
supersymmetric string vacua with the form
\begin{eqnarray}
\br^{d-1,1}\times \br_{\phi}\times \cM~,
\label{string vacua}
\end{eqnarray}  
where $\br_{\phi}$ denotes the $\cN=1$ Liouville theory 
($\cN=1$ linear dilaton SCFT) with the background 
charge $Q_{\phi}$ and $\cM$ is a rational $\cN=1$ SCFT with a central 
charge $c_{\cM}$. We would like to identify this conformal system 
as a string vacuum of singular special holonomy manifold realized as the
``cone over $\cM$'' in the decoupling limit $g_s\rightarrow 0$.
By assumption we have a linear dilaton background
$ \Phi(\phi)=-\frac{Q_{\phi}}{2}\phi$. The weak coupling region 
$\phi \sim +\infty$ 
corresponds to the holographic boundary $\br^{d-1,1}$ and 
the opposite region $\phi \sim -\infty $ is located around the singularity 
(``tip of the cone'').

The criticality condition is written as
\begin{eqnarray}
\frac{3}{2}(d-2)+\left(\frac{3}{2}+3Q_{\phi}^2\right)+c_{\cM}=12~,
\label{crit 1}
\end{eqnarray}
or equivalently,
\begin{equation}
\frac{Q_{\phi}^2}{8}=\frac{9-d}{16}-\frac{c_{\cM}}{24}~.
\label{crit 2}
\end{equation}

To build up the consistent string vacua the modular invariance is 
an important criterion. Let us first discuss the values of
Liouville momentum which enters into the modular invariant partition function.
We recall that the conformal dimension of a Liouville exponential $e^{\gamma\phi}$ is given 
by
\begin{equation}
h(e^{\gamma\phi})=-{1\over 2}\gamma^2-{1\over 2}Q_{\phi}\gamma=-{1\over 2}(\gamma+{Q_{\phi}\over 2})^2+{Q_{\phi}^2\over 8}
\end{equation}
It is known that the Liouville momentum in 
the partition function takes values in the
``principal continuous series" 
 (at least in the semi-classical analysis) \cite{Seiberg-L}
\begin{equation}
\gamma=-{Q_{\phi}\over 2}+ip, \hskip3mm p \in \br
\end{equation}
and thus 
\begin{equation}
h(e^{\gamma\phi})={1\over 2}p^2+{Q_{\phi}^2\over 8}\ge {Q_{\phi}^2\over 8}.
\end{equation}
Liouville exponential with momentum in the principal range represents a plane wave propagating in
spacetime and thus is a delta-function normalizable state. We note the presence of 
a ``mass" gap ${Q_{\phi}^2/8}$ in the Liouville spectrum.
Therefore, conformal blocks are  expanded only in terms of 
the ``massive representations'' of the extended superconformal
algebra which characterizes each special holonomy. 
The list of massive representations are given in appendix B.

~

We here summarize the aspects of spacetime SUSY in \eqn{string vacua} 
for various spacetime dimensions for later convenience.
\begin{itemize}
\item {\bf $SU(n)$-holonomy : } ~ 

This case corresponds to 
 $d=10-2n$, and the worldsheet SUSY
is required to be enhanced to $\cN=2$.
More explicitly, the criterion for
the $SU(n)$-holonomy is to ask whether it is possible to rewrite the theory
      as\footnote{In this paper we often use the concise expressions
         such as
$$
\cM \cong \frac{\cM}{H_k} \times H_k~,
$$
  where $H_k$ denotes the conformal theory of 
 the level $k$ $H$-current algebra.
  Precisely speaking, if $H$ is abelian, the R.H.S must be 
  interpreted as an orbifoldization with respect to the $H$-charge 
  as in the Gepner model \cite{Gepner}. 
   If $H$ is semi-simple, the R.H.S strictly
  means the ``projected tensor product'' discussed in \cite{Sotkov}.}
\begin{eqnarray}
\br_{\phi} \times \cM & \cong &
\left\lb \br_{\phi}\times U(1) \right\rb \times \frac{\cM}{U(1)}~,
\label{N=2 expression}
\end{eqnarray}
where $\br_{\phi}\times U(1)$ stands for the $\cN=2$ Liouville theory and 
the coset $\cM/U(1)$ describes an $\cN=2$ SCFT.  In the case when $\cM$
is an $\cN=1$ coset, $\cM/U(1)$ should be a Kazama-Suzuki model \cite{KS} 
associated to a K\"{a}hler homogeneous space.

When making  use of this criterion \eqn{N=2 expression},  
the most crucial point is as follows:
The  $\cN=2$ Liouville theory includes a compact boson $Y$
with the radius equal to $Q_{\phi}$
whose conformal blocks are written in terms of the theta functions 
of the level $n^2\cdot 2/Q_{\phi}^2$, where $n$ is an integer,
as discussed in \cite{ES1,Yamaguchi}. Ambiguity $n^2$ of the level
originates from the choice of possible momentum lattice of 
compact boson as shown  by the theta function identity \eqn{theta identity}.  

We have a natural geometric interpretation of \eqn{N=2 expression}:
The Calabi-Yau cones are build up over Sasaki-Einstein spaces,
which have the $U(1)$ fibration over the K\"{a}hler-Einstein manifolds
\begin{eqnarray}
\cM~\stackrel{U(1)}{\longrightarrow}~\cM/U(1)~.
\label{SE fibration}
\end{eqnarray}

The standard GSO projection with respect to the total $U(1)_R$-charge
of $\cN=2$ SCA leads to the spacetime SUSY as in Gepner models.
The explicit construction of supercharges is given in \cite{GKP} and 
we can check the cancellation of conformal blocks directly.

The characteristic features of the conformal blocks for 
the $SU(n)$-holonomy in each case $n=2,3,4$ is summarized as follows;
\begin{description}
 \item[(i) $SU(2)$-holonomy] : In this case the conformal blocks are expanded
       by the massive characters of $\cN=4$ superconformal algebra with
       $c=6$ (level 1), which have the forms as $q^{h-*}\th_3^2/\eta^3$
       in the NS sector \cite{ET}. The SUSY cancellation is expressed by 
       the familiar Jacobi's abstruse identity
\begin{eqnarray}
     \left(\frac{\th_3}{\eta}\right)^4-
     \left(\frac{\th_4}{\eta}\right)^4-\left(\frac{\th_2}{\eta}\right)^4
      \equiv 0 ~,
\label{Jacobi}
\end{eqnarray}
and we have 16 unbroken supercharges in 6-dimensional spacetime.
 \item[(ii) $SU(3)$-holonomy] : In this case the conformal blocks are expanded
       by the massive characters of ``$c=9$ extended superconformal
       algebra'' \cite{odake,EOTY}, 
      which have the form $q^{h-*}\Th{*}{1}\th_3/\eta^3$
       in the NS sector \cite{odake}. The SUSY cancellation is expressed as 
       the following identities
\begin{eqnarray}
&& \left\lb \left(\frac{\th_3}{\eta}\right)^2-
\left(\frac{\th_4}{\eta}\right)^2  \right\rb \frac{\Th{0}{1}}{\eta}
-\left(\frac{\th_2}{\eta}\right)^2 \frac{\Th{1}{1}}{\eta} \equiv 0~, 
\nonumber\\
&& \left\lb \left(\frac{\th_3}{\eta}\right)^2+
\left(\frac{\th_4}{\eta}\right)^2  \right\rb \frac{\Th{1}{1}}{\eta}
-\left(\frac{\th_2}{\eta}\right)^2 \frac{\Th{0}{1}}{\eta} \equiv 0~,
\label{cb cy3}
\end{eqnarray}
and we have 8 supercharges in 4-dimensional spacetime.
\item[(iii) $SU(4)$-holonomy] : In this case the conformal blocks are expanded
       by the massive characters of ``$c=12$ extended superconformal
       algebra'', which take the form $ q^{h-*}\Th{*}{3/2}\th_3/\eta^3$
       in the NS sector \cite{HS}. The SUSY cancellation is ensured by
       the identities
\begin{eqnarray}
&& 
\frac{\th_3}{\eta} \frac{\Th{0}{\frac{3}{2}}}{\eta} (\tau)
-\frac{\th_4}{\eta} \frac{\tTh{0}{\frac{3}{2}}}{\eta} (\tau)
-\frac{\th_2}{\eta} \frac{\Th{\frac{3}{2}}{\frac{3}{2}}}{\eta}(\tau)
\equiv 0, \nonumber \\
&&
\frac{\th_3}{\eta} \frac{\Th{1}{\frac{3}{2}}}{\eta} (\tau)
+\frac{\th_4}{\eta} \frac{\tTh{1}{\frac{3}{2}}}{\eta} (\tau)
-\frac{\th_2}{\eta} \frac{\Th{\frac{1}{2}}{\frac{3}{2}}}{\eta}(\tau)
 \equiv 0~,
\label{cb cy4}
\end{eqnarray}
and we have 4 supercharges in 2-dimensional spacetime.
\end{description}

~

\item {\bf $Sp(n)$-holonomy :}~

These superstring vacua correspond to $d=10-4n$ and are described by
the $\cN=4$ SCFT of $c=6n$ (level $n$). 
In the case when $\cM$ is defined as an $\cN=1$ supercoset CFT,
the condition for the $\cN=4$ enhancement of worldsheet SUSY 
has been studied in \cite{Sevrin}.
We have the (small) $\cN=4$ worldsheet SUSY, 
if and only if the following rewriting
is possible;
\begin{eqnarray}
\br_{\phi} \times \cM & \cong &
\left\lb \br_{\phi}\times SU(2) \right\rb \times \frac{\cM}{SU(2)}~,
\label{N=4 expression}
\end{eqnarray} 
where the $\cN=1$ coset $\cM/SU(2)$ is associated to 
a Wolf space (quaternionic symmetric space) \cite{Wolf}.

In the similar manner as \eqn{SE fibration}, the tri-Sasakian 
homogeneous spaces are known to have $SU(2)$-fibrations over Wolf spaces;
\begin{eqnarray}
\cM~\stackrel{SU(2)}{\longrightarrow}~\cM/SU(2)~,
\label{3S fibration}
\end{eqnarray}  
which yields the geometrical interpretation of \eqn{N=4 expression}.

Concerning the structure of conformal blocks,
the case of $Sp(1)(\cong SU(2))$-holonomy was already illustrated above.
In the case of $Sp(2)$-holonomy, on the other hand, 
the conformal blocks should be expanded by the massive characters
of $\cN=4$ SCA with $c=12$, which are written \cite{ET} as 
\begin{eqnarray}
 \frac{q^{h-*}}{\eta}\left(\frac{\th_3}{\eta}\right)^2 \chi_*^{SU(2)_1} 
  \equiv  \frac{q^{h-*}}{\eta}\left(\frac{\th_3}{\eta}\right)^2 
\frac{\Th{*}{1}}{\eta}~.
\label{mc sp(2)}
\end{eqnarray}
The  SUSY cancellation is expressed by the identity \eqn{cb cy3}
again. However, it turns out that we obtain  the $\cN=(3,3)$ (or $\cN=(6,0)$)
spacetime SUSY, that is, there exist only 6 supercharges in 2-dimensional spacetime.

~

\item {\bf $G_2$-holonomy :} ~

We set $d=3$, and the spacetime SUSY requires
the condition
\begin{eqnarray}
  U(1)_{3/2} \subset \cM~,
\label{cond non-compact G2}
\end{eqnarray}
where $U(1)_{3/2}$ denotes a $c=1$ CFT whose conformal blocks are described by  
characters $\Th{*}{3/2}(\tau)/\eta(\tau)$.
Precisely this relation means that the conformal blocks of the $\cM$-sector
are expanded in terms of characters 
$\Th{*}{3/2}(\tau)/\eta(\tau)$ and the
branching coefficients are not hit by the GSO projection.
The SUSY cancellation is realized again as \eqn{cb cy4}.  
If one takes account of the Liouville fermion, we have a relation \cite{SY1}
\begin{eqnarray}
 SO(1)_1 \times U(1)_{3/2} &\cong& \frac{SO(6)_1}{SU(3)_1} \times SO(1)_1 
\nonumber \\
&\cong& \frac{SO(7)_1}{(G_2)_1} \times \frac{(G_2)_1}{SU(3)_1} \nonumber \\
&\cong & \mbox{tri-critical Ising} \times \mbox{3-state Potts}~, 
\end{eqnarray}
Thus our condition is consistent 
with the criterion for the $G_2$-holonomy given in \cite{SV}.

~

 \item {\bf $Spin(7)$-holonomy : } ~

We set $d=2$, and the existence of 
spacetime SUSY requires the condition 
\begin{eqnarray}
  \mbox{tri-critical Ising} \subset \cM~,
\label{cond non-compact spin(7)}
\end{eqnarray}
which precisely means that the conformal blocks in the $\cM$-sector
can be decomposed by the $\cN=1$ characters of the tri-critical Ising
model (the first model in the $\cN=1$ minimal series with $c=7/10$).
The SUSY cancellation is realized as the following 
identities \footnote
    {In the convention here,
     the $\cN=1$ characters of tri-critical Ising are written 
    in terms of the $\cN=0$ ones as follows;
\begin{eqnarray}
&&
\chi^{\msc{tri}\,\sNS}_{0} = \chi^{\msc{tri}}_{0}+\chi^{\msc{tri}}_{3/2}~,~~~
\chi^{\msc{tri}\,\stNS}_{0} = \chi^{\msc{tri}}_{0}-\chi^{\msc{tri}}_{3/2}~,~~~
\chi^{\msc{tri}\,\sNS}_{1/10} = \chi^{\msc{tri}}_{1/10}
+\chi^{\msc{tri}}_{3/5}~,~~~
\chi^{\msc{tri}\,\stNS}_{1/10} 
= \chi^{\msc{tri}}_{1/10}-\chi^{\msc{tri}}_{3/5}~,
\nonumber \\
&& \chi^{\msc{tri}\,\sR}_{7/16} = 2 \chi^{\msc{tri}}_{7/16}~,~~~
\chi^{\msc{tri}\,\sR}_{3/80} = 2 \chi^{\msc{tri}}_{3/80}~.
\nonumber
\end{eqnarray}
};
\begin{eqnarray}
&&\sqrt{\frac{\th_3}{\eta}}\chi^{\msc{tri}\,\sNS}_{0}
-\sqrt{\frac{\th_4}{\eta}}\chi^{\msc{tri}\,\stNS}_{0}
-\sqrt{\frac{\th_2}{2\eta}}\chi^{\msc{tri}\,\sR}_{7/16}\equiv 0  \nonumber\\
&&\sqrt{\frac{\th_3}{\eta}}\chi^{\msc{tri}\,\sNS}_{1/10}
+\sqrt{\frac{\th_4}{\eta}}\chi^{\msc{tri}\,\stNS}_{1/10}
-\sqrt{\frac{\th_2}{2\eta}}\chi^{\msc{tri}\,\sR}_{3/80}\equiv 0 ~. 
\label{cb spin7}
\end{eqnarray}

\end{itemize}


~

\begin{table}[h]
 \begin{tabular}{|c|c|c|c|c|}\hline
 dim & name & worldsheet SUSY & algebra 
                               & structure of massive 
 reps. \\\hline\hline
 4 & hyper K\"ahler & $\cN=4$ & $SU(2)_1$ & $SO(4)_1$ \\ \hline
 6 &Calabi-Yau & $\cN=2$ & $U(1)_{3/2}$ & $SO(2)_1\times SU(2)_1$  \\ \hline
 7 & $G_2$ & $\cN=1$  & tri-critical Ising 
   & $SO(1)_1 \times U(1)_{3/2}$ \\ \hline
 8 & hyper K\"ahler & $\cN=4$ & $SU(2)_2$ & 
   $SO(4)_1 \times SU(2)_1$ \\ \hline
 8 &Calabi-Yau & $\cN=2$ & $U(1)_2$ & $SO(2)_1\times U(1)_{3/2}$\\ \hline
 8 & $Spin(7)$ & $\cN=1$ & Ising &  $SO(1)_1\times$ 
 tri-critical Ising\\ \hline
 \end{tabular}
\caption{We summarize the algebraic structures of worldsheet theories 
describing manifolds with special holonomies. The ``structure for
the massive reps.'' is the algebraic structure which is manifest in
the characters of massive representations.}
\end{table}


~
\subsection{Constructions of Spacetime Supercharges}


Although we have just seen the characteristic 
features of conformal blocks in the string vacua 
\eqn{string vacua}, it may still be helpful to construct explicitly
the spacetime supercharges for these vacua.
We shall only consider the left-movers for simplicity. It is straightforward 
to combine the left and right movers so as to be consistent with the 
GSO conditions for the type IIA or type IIB string theories.

~

\noindent
{\bf 1. Supercharges for the $SU(n)$-holonomies : }

Construction of supercharges in these cases 
is quite standard and presented in \cite{GKP}.
We prepare the spin fields $S_0^{\al\,(\pm)}$ for the Minkowski 
spacetime $\br^{d-1,1}$ with the conformal weight
$h=d/16$, where $\al=1,2,\ldots, 
2^{\left\lb\frac{d-2}{2}\right\rb}$, and $\pm$ denotes 
the chirality. We also introduce the spin fields 
for the sector of internal space described by the $\cN=2$ SCFT
\begin{eqnarray}
\br_{\phi} \times \cM \cong 
\lb \br_{\phi} \times U(1) \rb \times \frac{\cM}{U(1)}~.
\end{eqnarray}
The total $U(1)_R$ current $J_R$ is bosonized as 
\begin{eqnarray}
J_R = i\partial H~, ~~~ H(z)H(0) \sim -n\ln z~,
\end{eqnarray}
and the relevant spin fields are defined as 
\begin{eqnarray}
S^{\pm} = e^{\pm \frac{i}{2}H} ~,
\end{eqnarray}
which have the conformal weight $h=n/8$.

The GSO projection gives the chiral supercharges for $d=2,6$, 
and non-chiral supercharges for $d=4$;
\begin{eqnarray}
 d=2,6~&& :~~~
  Q_{\al}^{\pm} = \oint S_0^{\al\, (+)} S^{\pm} e^{-\frac{\varphi}{2}}~,
\label{sc cy 1}
\\
 d=4~ && : ~~~
  Q_{\al}^{+} = \oint S_0^{\al\, (+)} S^{+} e^{-\frac{\varphi}{2}}~,~~~
  Q_{\dal}^{-} = \oint S_0^{\dal\, (-)} S^{-} e^{-\frac{\varphi}{2}}~,
\label{sc cy 2}
\end{eqnarray}
where $\varphi$ denotes the standard bosonized superghost.

~

\noindent
{\bf 2. Supercharges for the $Spin(7)$-holonomy}

The $Spin(7)$-holonomy admits one Killing spinor with a definite 
chirality. We thus assume the spin-field in $\br^{1,1}$ should 
have a definite chirality, say, $S^{(+)}_0$. \footnote
      {Without this assumption, we would have twice as many supercharges
       as those expected from supergravity. In fact, we further construct  
        the BRST invariant supercharge 
       $\dsp \tilde{Q} = \oint S^{(-)}_0\left(\sigma^{\phi}_+\sigma^{\cM}_-
+\sigma^{\phi}_-\sigma^{\cM}_+\right) e^{-\varphi/2}$ 
($\dsp \tilde{Q}' = \oint S^{(-)}_0\left(\sigma^{\phi}_+\sigma^{\cM}_+
+\sigma^{\phi}_-\sigma^{\cM}_-\right) e^{-\varphi/2}$) which is mutually
local with $Q$ ($Q'$) \eqn{sc spin(7)}. 
  This disagreement is probably an artifact originating from the fact 
  that we are now working on the singular background without Liouville potential  
   terms, while the 2-dimensional supergravity should correspond to 
   low energy effective theory on smooth backgrounds. }   
In order to define the spin fields for the internal space, 
we treat the Liouville and $\cM$ sectors separately. 
Since the $\cN=1$ Liouville sector includes a free fermion,
we have the doubly degenerate spin fields $\sigma_{\pm}^{\phi}$,
which have the conformal weight $h=1/16$ and satisfy the OPEs 
\begin{eqnarray}
&&\hskip-3mm  \psi^{\phi}(z)\sigma_{\pm}^{\phi}(0) \sim \frac{\pm i}{\sqrt{2}z^{1/2}} 
\sigma^{\phi}_{\mp} ~, ~~
\sigma_{\pm}^{\phi}(z)\sigma_{\pm}^{\phi}(0)
\sim \frac{1}{z^{1/8}}~, ~~
\sigma_{+}^{\phi}(z)\sigma_{-}^{\phi}(0) \sim 
\frac{z^{3/8}}{\sqrt{2}} \psi^{\phi}(0).
\label{Liouville spin}
\end{eqnarray}
On the other hand, the assumption \eqn{cond non-compact spin(7)}
implies the existence of spin fields from tri-critical Ising model with $h=7/16$.
They are again doubly degenerate and  we denote them as
$\sigma^{\cM}_{\pm}$.

The candidate supercharges are now written as
$\dsp \sim \oint S^{(+)}_0 \sigma^{\phi}_*\sigma^{\cM}_* 
e^{-\varphi/2}$. To achieve the correct construction
we must take account of the BRST invariance, especially the condition 
$G_0(\equiv G^{\phi}_0+G^{\cM}_0)=0$. 
The following relation is helpful for this purpose;
\begin{eqnarray}
G^{\cM}(z)\sigma_{\pm}^{\cM}(0) &\sim & \frac{1}{z^{3/2}}\,
\sqrt{\frac{7}{16}-\frac{c_{\cM}}{24}} \sigma_{\mp}^{\cM}(0)  \nonumber \\
&=& \frac{1}{z^{3/2}} \frac{Q_{\phi}}{2\sqrt{2}}\sigma_{\mp}^{\cM}(0)~,
\label{OPE sigma M}
\end{eqnarray}
where the first line is deduced from $ (G^{\cM}_0)^2 = L_0^{\cM}
-\frac{c_{\cM}}{24}$, and the second line is due to \eqn{crit 2}.
Now, it is not difficult to find out the following BRST invariant
combinations;
\begin{eqnarray}
Q= \oint S^{(+)}_0\left(\sigma^{\phi}_+\sigma^{\cM}_+
+\sigma^{\phi}_-\sigma^{\cM}_-\right) e^{-\varphi/2}~,~~~
Q'= \oint S^{(+)}_0\left(\sigma^{\phi}_+\sigma^{\cM}_-
+\sigma^{\phi}_-\sigma^{\cM}_+\right) e^{-\varphi/2}~.
\label{sc spin(7)}
\end{eqnarray}
They are mutually non-local and we must take only one of them,
which amounts to process of GSO projection. We have thus obtained 
the desired supercharge.

~

\noindent
{\bf 3. Supercharges for the $G_2$-holonomy}

Construction of supercharges for the $G_2$-holonomy 
is almost parallel. The condition \eqn{cond non-compact G2}
ensures the existence of the doubly degenerate spin fields
$\sigma^{\cM}_{\pm}$ with $h=3/8$, and the BRST invariant 
supercharges are found out to be 
\begin{eqnarray}
Q^{\al}= \oint S^{\al}_0\left(\sigma^{\phi}_+\sigma^{\cM}_+
+\sigma^{\phi}_-\sigma^{\cM}_-\right) e^{-\varphi/2}~,~~~
Q^{'\,\al}= \oint S^{\al}_0\left(\sigma^{\phi}_+\sigma^{\cM}_-
+\sigma^{\phi}_-\sigma^{\cM}_+\right) e^{-\varphi/2}~,
\label{sc G2}
\end{eqnarray}
where $S^{\al}_0$ ($\al=1,2$) is the 3-dimensional spin field with 
the conformal weight $h=3/16$.
$Q^{\al}$ and $Q^{'\,\al}$ are again mutually non-local, and
we have to take either one by the GSO projection.

~

\noindent
{\bf 4. Supercharges for the $Sp(2)$-holonomy}

The last case is the most non-trivial.
The $Sp(2)$-holonomy admits three independent 
Killing spinors with the same chirality.
So, we assume the longitudinal part 
of spin fields is $S^{(+)}_0$,
as in the case of $Spin(7)$-holonomy.

To define the spin fields in the internal space, 
we first recall the $SU(2)_2$-current algebra $\{K^a(z)\}$
describing the $SU(2)_R$-symmetry in the $\cN=4$ superconformal
theory, and the total $U(1)_R$ current of the $\cN=2$ SCFT
is identified as $2K^3(z)$. Moreover, 
since all conformal blocks are expanded into massive characters \eqn{mc sp(2)}, 
the $SU(2)_2$ current algebra is also embedded in $SU(2)_1\times
SO(4)_1$. 
In particular, we have the following decomposition of the total $U(1)_R$-current;
\begin{eqnarray}
&& 2K^3(z)=i\partial H_1(z) +i\partial H_2(z)+\sqrt{2}i\partial H_3(z)~, \\
&& H_i(z)H_j(0) \sim -\delta_{ij}\ln z~.
\end{eqnarray} 
Here the compact bosons $H_1$, $H_2$ have radii equal to 1, and 
describe 1+3=4 free fermions corresponding to  
the superpartners of the Liouville mode and the  
$SU(2)$-factor in \eqn{N=4 expression} (the fermions along the fiber of
\eqn{3S fibration}). 
$H_3$ has the radius  equal to $\sqrt{2}$ (self-dual radius) and 
generates the $SU(2)_1$-current algebra.
We now introduce the following spin fields (up to cocycle factors)
\begin{eqnarray}
S^{\ep_1\ep_2\ep_3} = e^{i\frac{\ep_1}{2}H_1 +i\frac{\ep_2}{2}H_2+
    i\frac{\sqrt{2}\ep_3}{2}H_3}~,~~~(\ep_i =\pm 1)~.
\label{spin sp(2)}
\end{eqnarray}
Candidate supercharges are now (linear combinations of) 
$\dsp \oint S^{(+)}_0S^{\ep_1\ep_2\ep_3}e^{-\varphi/2}$ which 
have 8 independent components.
Again the non-trivial condition for the BRST invariance is the constraint
$G_0 \equiv G^{\cM}_0+G^{\phi}_0 =0$, and one finds the following solutions
\begin{eqnarray}
&& Q^{\pm} = \oint S^{(+)}_0S^{\pm\pm\pm}e^{-\varphi/2}~, \nonumber \\
&& Q^3 = \frac{1}{2}\oint S^{(+)}_0\left(S^{++-}+S^{--+}\right)e^{-\varphi/2}~,
\label{sc sp(2) t}
\end{eqnarray}
and also
\begin{eqnarray}
Q' = \frac{1}{2}\oint S^{(+)}_0\left(S^{+-+}-S^{-+-}\right)e^{-\varphi/2}~.
\label{sc sp(2) s}
\end{eqnarray}
As is easily shown, $Q^+$, $Q^-$ are the same supercharges as 
those of $SU(4)$-holonomy \eqn{sc cy 1}. Moreover,  
$Q^a (a=\pm,3)$ composes a triplet of $SU(2)_R$ and $Q'$ is a singlet.
In particular,  $\{Q^a\}$ are identified as the spin 1 primary fields of 
the current algebra $K^a(z) (a=\pm,3)$.  
$Q^a$ and $Q'$ are mutually non-local, 
and we should choose either one by the GSO projection.
We take the triplet $Q^a$ to recover the string vacua of 
$Sp(2)$-holonomy.

~

Finally let us make a comment.
It is well-known \cite{KW,AFHS,Gubser,MP} that the Ricci flat background
$\br^{d-1,1} \times C(X_{9-d})$ is converted into $AdS_{d+1} \times X_{9-d}$
by letting $(d-1)$-branes ``fill''  the Minkowski spacetime 
$\br^{d-1,1}$, and taking the near horizon limit.
We here denote the $(9-d)$-dim. Einstein space as $X_{9-d}$ and 
its Ricci flat cone as $C(X_{9-d})$. Even though the worldsheet approach 
to string theory on such backgrounds is usually very difficult due to 
the RR-flux, the $d=2$ cases are tractable. 
Namely, we can fill the NS1 branes and interpolate 
the background $\br^{1,1}\times C(X_7)$
to $AdS_3 \times X_7$. 

The analogous relation in our study of ``cone
over SCFT'' could be depicted as 
\begin{eqnarray}
 \br^{1,1} \times \br_{\phi} \times \cM ~ 
\stackrel{+\msc{NS1}}{\Longrightarrow} ~
 AdS_3 \times \cM~.
\label{interpolation}
\end{eqnarray}
In the R.H.S, the $AdS_3$ sector is described by the $SL(2;\br)$ super
WZW model with the (bosonic) level $k$ which is determined by the relation
$ Q_{\phi}^2 = \frac{2}{k+2}$. 
This relation \eqn{interpolation} has been already pointed 
out in \cite{GKP} in the cases of $SU(4)$-holonomy. There it is also discussed 
that the spacetime SUSY algebra should be enhanced to the 
$\cN=2$ superconformal algebra (only for the left or right movers)
acting on the boundary of $AdS_3$, which is explicitly constructed 
in \cite{GRBL} along the same line as \cite{GKS}.
The similar observations are also  possible for the $Spin(7)$ and
$Sp(2)$-holonomy cases:
\begin{description}
 \item[(i) $Spin(7)$-holonomy : ]
The easiest way to move from the L.H.S to R.H.S in 
\eqn{interpolation} is by making the formal replacements;
$\psi^0\,\rightarrow\,i\Psi^2$, $\psi^1\,\rightarrow\, \Psi^1$,
and $\psi^{\phi}\,\rightarrow\,i\Psi^3$, where $\psi^0$, $\psi^1$,
$\psi^{\phi}$ are the free fermions along the longitudinal and 
Liouville directions and $\Psi^A$ denotes the fermionic coordinates in 
the $\cN=1$ $SL(2;\br)$-WZW model. 
The structure of spin fields are quite similar.
The essential difference lies in the BRST charges, especially in 
the definitions of superconformal currents $G(z)$: 
$G(z)$ of $SL(2;\br)$ super WZW model includes a cubic fermionic
term $\sim \Psi^1\Psi^2\Psi^3$, while that of $\cN=1$ Liouville 
theory ($\times \br^{1,1}$) only includes a linear term 
$\sim Q_{\phi} i \partial \psi^{\phi}$.

With these preparations, we can construct twice as many 
supercharges for the $AdS_3\times \cM$ backgrounds as compared with 
\eqn{sc spin(7)};
\begin{eqnarray}
&& \cG_{+1/2}= \oint S^{(+)}_0\left(\sigma^{3}_+\sigma^{\cM}_+
+\sigma^{3}_-\sigma^{\cM}_-\right) e^{-\varphi/2}~, \nonumber \\
&& \cG_{-1/2}= \oint S^{(-)}_0\left(\sigma^{3}_+\sigma^{\cM}_-
- \sigma^{3}_-\sigma^{\cM}_+\right) e^{-\varphi/2}~,
\label{sc spin(7) 2}
\end{eqnarray}
where we denote the spin fields associated $\Psi^1$, $\Psi^2$ 
as $S^{(+)}_0$, and $\sigma_{\pm}^3$ is defined similarly as in 
\eqn{Liouville spin} with respect to $\Psi^3$. 
They are naturally identified as the ``zero-modes'' of the  
spacetime $\cN=1$ superconformal algebra.  It is also not difficult
to construct the full superconformal current oscillators 
based on the Wakimoto free field representation along the line 
of \cite{GKS}. 
\item[(ii) $Sp(2)$-holonomy : ]
For the $Sp(2)$-holonomy the argument is almost parallel.
In the background $AdS_3 \times \cM$ we can obtain twice as many
supercharges
\begin{eqnarray}
&& \cG_{+1/2}^{\pm} = 
\oint S^{(+)}_0S^{\pm\pm\pm}e^{-\varphi/2}~, \nonumber \\
&& \cG_{+1/2}^3 
= \frac{1}{2}\oint S^{(+)}_0\left(S^{++-}+S^{--+}\right)e^{-\varphi/2}~,
\nonumber \\
&& \cG_{-1/2}^{\pm} = 
\oint S^{(-)}_0S^{\mp\pm\pm}e^{-\varphi/2}~, \nonumber \\
&& \cG_{-1/2}^3 
= \frac{1}{2}\oint S^{(-)}_0\left(S^{-+-}+S^{+-+}\right)e^{-\varphi/2}~.
\label{sc sp(2) 2}
\end{eqnarray}
They are again enhanced to the full generators of $\cN=3$ superconformal
algebra \cite{SS}, where the superconformal currents 
behave as  a triplet of $SU(2)_R$ symmetry,
and coincide with those constructed  in \cite{AGS}
(up to normalizations and the convention of spin fields). 
In addition, we will show in the next section 
that the $Sp(2)$-holonomy can be achieved only
for $\cM=SO(5)/SO(3)$, $SU(3)/U(1)$,  if we assume supercoset CFT's 
for the $\cM$-sector.  This fact is consistent with the observation
given in \cite{AGS}.
\end{description}

~

\section{Supercoset CFT's for Superstring Vacua of Special Holonomies}

\subsection{Preliminary: Some Notes on Supercoset CFT's}

In order to fix our discussions we shall adopt 
the coset construction for the $\cM$-sector from now on. 
The $\cN=1$ supercoset CFT's are easily constructed  by using
the super affine Kac-Moody algebras. They have the general form
\begin{eqnarray}
  \cM =  \frac{G_k\times SO(D)_1}{H}~, ~~~\dim G/H=D~,
\label{N=1 coset}
\end{eqnarray}
with a Lie group $G$ and its subgroup $H$. $SO(D)_1$ stands for the
current algebra generated by $D$ free fermions. The condition
$\dim G/H=D$ ensures the ${\cal N}=1$ superconformal symmetry. 
We especially concentrate on the cases of $D=9-d$ for 
the superstring vacua of the type $\br^{d-1,1}\times \br_{\phi}\times
\cM$ (namely, for the $10-d$ dimensional internal space).
The reason why we do so is as follows: 
The coset spaces $G/H$ are generically endowed with Einstein metrics 
and were studied extensively in old days of Kaluza-Klein supergravity 
\cite{Witten-G,DFN,CR,PP,CDF,CRW,Romans,DNP}.
The cones over these spaces are known to possess
Ricci flat metrics \cite{KW,AFHS,Gubser,MP}, leading to supersymmetric
string vacua, when the coset spaces possess the geometrical structures 
listed in the table 1. 
In a formal sense the string vacua \eqn{string vacua} we are studying 
are the CFT versions of the Ricci flat cones.
We point out that $ c_{\cM} \leq 3D/2$
holds in general (equality holds at $k=\infty$). 
Thus we always have the real background charge
$Q_{\phi}$. (Recall the condition \eqn{crit 2}.)

The level of the current algebra of $H$ is slightly non-trivial 
to determine and depends on the embedding of $H$ in general.
We assume the decomposition $H=H_0\times H_1 \times \cdots \times H_r$,
where $H_0$ is the abelian part and $H_i$ ($i=1,\ldots,r$) are simple
parts. 
The level $k_i$ of each simple factor $H_i$ is defined in the standard manner;
\begin{eqnarray}
J_{(i)}^{a}(z)J_{(i)}^{b}(0) &\sim& 
\frac{k_i(t_{(i)}^a,t^b_{(i)})_i}{z^2}
   +\frac{if_{(i)\,~~c}^{\,\, ~ab}}{z}J^c_{(i)}(0)~~~(\mbox{for $i\neq 0$})~,
\nonumber \\
J_{(0)}^{a}(z)J_{(0)}^{b}(0) &\sim& 
\frac{k_0(t_{(0)}^a,t^b_{(0)})}{z^2}~~~(\mbox{for $H_0$})~.
\end{eqnarray} 
where $\{t_{(i)}^a\}$ is a basis of the Lie algebra of
$H_i$ and $J_{(i)}^a(z)\equiv (J_{(i)}(z), t_{(i)}^a)_i$. 
The Killing metrics $(~,~)$ for $G$ and 
$(~,~)_i$ for the $H_i$ sectors  are
canonically normalized as $(\th,\th)=2$, 
$(\theta_i,\theta_i)_i=2$, where $\th$,  $\theta_i$ 
are the highest roots of $G$, $H_i$, respectively. 
Notice that the inner products $(~,~)$ and $(~,~)_i$ have different
normalizations in general.  $(\th_i,\th_i)$ is not 
necessarily equal to 2 and depends on the choice of embedding of $H_i$.
Now, the levels $k_i$ are determined by comparing the Schwinger terms 
of the {\em super} affine Lie algebras of $G$ and $H_i$;  
\begin{eqnarray}
&& k_0=k+g^*   \nonumber \\
&& k_i=\frac{2}{(\th_i,\th_i)}(k+g^*)-h^*_i~,~~~(i=1,\ldots,r)~, 
\label{level formula}
\end{eqnarray}
where $g^*$, $h^*_i$ denote the dual Coxeter numbers of $G$, $H_i$
respectively\footnote
     {In the case $H_i=SO(3)$, one must use the formula
$$
 k_i=\frac{1}{(\th_i,\th_i)}(k+g^*)-1
$$ 
instead of \eqn{level formula}. This fact is due to the
equivalence $SO(3)_k \cong SU(2)_{2k}$ as an affine Lie algebra.}.
Care is needed when $G$ is non-simply laced 
and some $H_i$'s are embedded along its short roots.
The central charges of these coset CFT's
are evaluated as
\begin{eqnarray}
 c_{\Mcal}&=& \frac{k \dim G}{k+\hc}+\frac{1}{2}D -
\sum_{i=0}^r \frac{k_i \dim H_i}{k_i+\hs_i} ~.
\label{c M}
\end{eqnarray}

We also remark that the sub-root lattice describing the charge spectrum 
of $H_0$-sector must be specified in order to define the coset model uniquely.
The conformal blocks of this sector are written in terms of  the 
theta functions associated to the charge lattice 
$\Gamma \subset \sqrt{k+g^*} Q$, where $Q$ is the root lattice of $G$. 
They are explicitly written as 
\begin{eqnarray}
F^{H_0}_{\la}(\tau) &=& 
\frac{\Theta^{(\Gamma)}_{\la}(\tau)}{\eta(\tau)^L}~,~~~(\dim H_0=L)
\nonumber \\
\Theta^{(\Gamma)}_{\la}(\tau)&= &
\sum_{\al \in \la+\Gamma}\,q^{\frac{1}{2}(\al,\al)} ~,
~~~(\la \in \Gamma^*)~.
\end{eqnarray} 
For example, suppose that the charge lattice is given by
\begin{eqnarray}
&&\Gamma = \bz \nu_1 + \cdots +\bz \nu_L~,~~~(L=\dim H_0)~, \nonumber \\
&& (\nu_{a},\nu_{b})=0~,~~~({}^{\forall} a\neq b)  ~,~~~ 
 (\nu_a,\nu_a) = 2l_a(k+g^*)~,
\end{eqnarray} 
then we have the decomposition of the theta functions 
\begin{eqnarray}
&& \Theta^{(\Gamma)}_{\la}(\tau)
= \prod_{a=1}^L \Th{m_a}{l_a(k+g^*)}(\tau)~, ~~~
  \la = \sum_{a=1}^L \, m_a\nu_a^*   ~, 
\end{eqnarray} 
where $\nu_a^*$ are the dual bases of $\Gamma^*$ such that 
$(\nu_a,\nu_b^*)=\delta_{ab}$.
In this paper we shall adopt the convention that ``$U(1)_k$'' means
the $c=1$ conformal theory composed of the conformal blocks of the forms
$\Theta_{*,k}(\tau)/\eta(\tau)$. 
Namely, we here find 
\begin{eqnarray}
 (H_0)_{k+g^*}\cong U(1)_{l_1(k+g^*)} \times \cdots \times U(1)_{l_L(k+g^*)}
\end{eqnarray}
We will later face non-trivial 
examples in which the different choice of charge lattice leads
to inequivalent string vacua.

~


\subsection{Coset Constructions of $d=6$ Superstring Vacua}

We start with the study of $d=6$ string vacua.
The spacetime SUSY corresponds  to the $SU(2)$-holonomy in this case.
Assuming the $\cN=1$ supercoset CFT $\cM =G/H$ with the condition that 
$G$ is compact, simple and $\dim G/H=3$,
the possible example is
$G=SU(2)$, $H=\{\mbox{id}\}$.
We also study the case $G=SO(4)$, $H=SO(3)$ as a paticular example of
a semi-simple $G$.

The first example, $G=SU(2)$, $H=\{\mbox{id}\}$, 
is the familiar CHS $\sigma$-model describing the NS 5-brane \cite{CHS};
\begin{eqnarray}
\br_{\phi} \times \psi^{\phi} \times \cM &=&
\br_{\phi} \times \psi^{\phi} \times SU(2)_k \times SO(3)_1 \nonumber \\
& \cong & \left\lb \br_{\phi} \times \psi^{\phi} \times 
U(1)_{k+2} \times SO(1)_1 \right\rb \times 
\frac{SU(2)_k\times SO(2)_1}{U(1)_{k+2}}~.
\label{chs}\end{eqnarray} 
Liouville field, together with the WZW model $SU(2)_k$, describes the configuration of the
throat region ${\bf R}_+\times S^3$ transverse to the NS 5-brane.
In the last line the part described by $\lb \cdots \rb$ is 
the $\cN=2$ Liouville theory with 
$ Q_{\phi}=\sqrt{2/(k+2)}$ and the remaining part is the 
$\cN=2$ minimal model of the level $k$.  
In \cite{OV} this system is interpreted as 
describing the string theory compactified on 
an ALE space
with the $A_{k+1}$-type singularity
(The $D$ and $E$-type singularities  are naturally incorporated 
as the modular data of $SU(2)_k$ sector \cite{ABKS}.).
Thus (\ref{chs}) represents the well-known 
T-duality between NS 5-brane and ALE space \cite{OV,GHM}.

In the second example, $G=SO(4)$, $H=SO(3)$, 
we have two possibilities of $SO(3)$-embedding.
The first choice is the embedding into one of the $SO(3)$'s of 
$SO(4) \simeq SO(3)\times SO(3)$, which again reduces 
to the CHS $\sigma$-model because
of the relation
\begin{eqnarray}
\cM&=& \frac{SO(4)_k\times SO(3)_1}{SO(3)_{k/2}} \nonumber \\
&\cong& SU(2)_k \times SO(3)_1 ~.
\label{SO(4)-SO(3) 1}
\end{eqnarray}
The second choice is the diagonal embedding in $SO(3)\times SO(3)$,
which leads to
\begin{eqnarray}
\cM&=& \frac{SO(4)_k\times SO(3)_1}{SO(3)_{k+1}} ~.
\label{SO(4)-SO(3) 2}
\end{eqnarray}
This case corresponds to vacua with no spacetime SUSY.

In addition, we can consider the following generalization of
\eqn{SO(4)-SO(3) 2}
\begin{eqnarray}
\cM&=& \frac{SU(2)_{k_1}\times SU(2)_{k_2} \times SO(3)_1}
{SU(2)_{k_1+k_2+2}}~.
\label{SU(2) diagonal}
\end{eqnarray}
These $\cN=1$ diagonal coset theories have been studied 
intensively in \cite{Noyvert}.
It is obvious that \eqn{SU(2) diagonal} cannot provide 
any $d=6$ supersymmetric vacuum. However, we will later show
that the $d=3$ and $d=2$ supersymmetric vacua of this form
exist under the suitable restrictions of the levels
of current algebras.

~

\subsection{Coset Constructions of $d=4$ Superstring Vacua}

In this case the supersymmetric string vacua correspond 
to $SU(3)$-holonomy. The criterion for spacetime SUSY
is thus whether the worldsheet SUSY is enhanced to $\cN=2$.

We assume $\dim G/H =5$. 
If $G$ is simple, we only have the possibilities 
\begin{equation}
G/H= SO(6)/SO(5)~,~SU(3)/SU(2)~.
\end{equation}
We also study the case $G/H=(SU(2)\times SU(2))/U(1)$ 
as a particular example of a semi-simple $G$.


\noindent
{\bf 1. $ ~ G/H=SO(6)/SO(5)$ }

In this case the $\cN=1$ supercoset CFT is given as
\begin{eqnarray}
\cM = \frac{SO(6)_k\times SO(5)_1}{SO(5)_{k+1}}~.
\end{eqnarray}
The worldsheet SUSY for the total system 
$\br_{\phi} \times \psi^{\phi}\times \cM$ 
is not enhanced since the coset $\cM/U(1)$ is not well-defined 
in this case. Therefore, this model corresponds 
to a string vacuum with no spacetime SUSY.

~


\noindent
{\bf 2. $ ~ G/H=SU(3)/SU(2)$}


We have two possibilities of the embedding of $SU(2)$.

\begin{description}
 \item[(i) $SU(2)$ embedded as the ``isospin subgroup'' : ]

~

The easier embedding is of course into
the usual isospin subgroup of $SU(3)$.
In this case, we can show that the worldsheet SUSY is enhanced to $\cN=2$
because of the equivalence
\begin{eqnarray}
\br_{\phi}\times \psi^{\phi}\times \cM&\cong&
\lb \br_{\phi}\times \psi^{\phi} \times U(1)_{3(k+3)}\times SO(1)_1\rb
\times \frac{SU(3)_k\times SO(4)_1}{SU(2)_{k+1}\times U(1)_{3(k+3)}}~,
\nonumber \\
&& \hspace{10cm}
\label{rewriting S5}
\end{eqnarray} 
as we already mentioned in \eqn{N=2 expression}. 
The part $\lb \cdots \rb$ is interpreted as the $\cN=2$ Liouville. 
In fact, the criticality condition \eqn{crit 2} gives us 
$ Q_{\phi}^2 = 6/(k+3)$, resulting in $3^2\cdot 2/Q_{\phi}^2 = 3(k+3)$.
Thus,  the $U(1)_{3(k+3)}$ piece is identified as the compact boson 
of $\cN=2$ Liouville theory.
The remaining coset CFT is the Kazama-Suzuki model 
for $\CP_2$. In this way we have confirmed that  
this string vacuum corresponds to a non-compact $CY_3$ manifold.

Notice that the coset $SU(3)/SU(2)$ is isomorphic with $S^5$,
which is an elementary example of Sasaki-Einstein manifold
with the $U(1)$-fibration
\begin{eqnarray}
S^5~\fiber{U(1)}~\CP_2~.
\end{eqnarray}
This is the geometrical interpretation of \eqn{rewriting S5}.


\item[(ii) $SU(2)$ embedded as the maximal subgroup : ]

~

Another embedding is defined 
by identifying the simple root of $SU(2)$ with $\theta/2$,
where $\th\equiv \al_1+\al_2$ is the highest root of $SU(3)$.
In practice, this is given by restricting the canonical action of 
$SU(3)$ on complex 3-vectors to that of $SO(3) (\simeq SU(2))$ 
on real 3-vectors. By this embedding,
the adjoint representation of $SU(3)$
is decomposed as ${\bf 8\,\rightarrow\,3+5}$, and thus it is ``maximal''
(see \cite{CRW} for the detail). The coset space is again 
isomorphic with $S^5$. 

Since $(\theta/2)^2=1/2$ holds, we find that  
the relevant supercoset CFT should have the following form due to the formula 
\eqn{level formula}
\begin{eqnarray}
\cM&=& \frac{SU(3)_{k}\times SO(5)_1}
{SU(2)_{4k+10}} ~.
\label{SU(3)-SU(2) max}
\end{eqnarray}
It may be helpful to confirm that this coset CFT is really well-defined.
To this aim it will be enough to check the existence of $SO(5)_1/SU(2)_{10}$.
In fact, this coset CFT corresponds to the maximal embedding 
$SU(2) \subset SO(5)$ which we later consider. 
Another explanation is given as follows:
We have a conformal embedding $SU(2)_{10} \subset SO(5)_1$
($c=5/2$ for both of $SO(5)_1$, $SU(2)_{10}$) based on 
the $E_6$-type modular invariance of $SU(2)_{10}$ \cite{CIZ}.
Namely,  the corresponding partition function is given by
\begin{eqnarray}
Z=\left|\chi^{(10)}_0+\chi^{(10)}_6\right|^2
+\left|\chi^{(10)}_3+\chi^{(10)}_7\right|^2
+\left|\chi^{(10)}_4+\chi^{(10)}_{10}\right|^2~,
\end{eqnarray} 
where we denote the $SU(2)_k$ character of spin $\ell/2$ as 
$\chi^{(k)}_{\ell}$, 
and the following character relations are known 
(see \cite{CFTtext}, for example);
\begin{eqnarray}
&& \chi^{SO(5)_1}_{\msc{\bf basic}}(\tau)
=\chi^{(10)}_0(\tau)+\chi^{(10)}_6(\tau)~,~~~
\chi^{SO(5)_1}_{\msc{\bf spinor}}(\tau)
=\chi^{(10)}_3(\tau)+\chi^{(10)}_7(\tau)~, \nonumber\\
&& \chi^{SO(5)_1}_{\msc{\bf vector}}(\tau)
=\chi^{(10)}_4(\tau)+\chi^{(10)}_{10}(\tau)~.
\end{eqnarray}

We have no extra $U(1)$ symmetry in \eqn{SU(3)-SU(2) max}
since the $SU(2)$ is a maximal subgroup, and thus
the worldsheet SUSY cannot be enhanced. 
Therefore, the corresponding string vacua are not supersymmetric.

\end{description}

~


\noindent
{\bf 3. $ ~ G/H=(SU(2)\times SU(2))/U(1) $ }

Lastly, let us consider the most non-trivial example 
\begin{eqnarray}
\cM=\frac{SU(2)_{k_1}\times SU(2)_{k_2}\times SO(5)_1}{U(1)}~.
\end{eqnarray}
As was already illustrated,
the charge lattice of $U(1)$ must have the following form
because of the worldsheet SUSY
\begin{eqnarray}
\Gamma = \bz\left(p\sqrt{k_1+2}\al+q\sqrt{k_2+2}\beta\right)~,
\label{gamma Tpq}
\end{eqnarray}
where $\al$, $\beta$ denote the simple roots of the two $SU(2)$ factors
and $p, q\in \bz_{\geq 0}$ ($(p,q)\neq (0,0)$).
We can reexpress $\cM$ as 
\begin{eqnarray}
\br_{\phi}\times \psi^{\phi} \times \cM & \cong &
\br_{\phi}\times \psi^{\phi} \times 
\frac{SU(2)_{k_1}\times SO(2)_1}{U(1)_{k_1+2}} \times 
\frac{SU(2)_{k_2}\times SO(2)_1}{U(1)_{k_2+2}} \nonumber \\
&& \hspace{1in} \times \frac{U(1)_{k_1+2}\times U(1)_{k_2+2}}
{U(1)_{p^2(k_1+2)+q^2(k_2+2)}} \times SO(1)_1 \nonumber \\
&\cong & \lb \br_{\phi}\times \psi^{\phi} 
\times U(1)_{(k_1+2)(k_2+2)\left\{p^2(k_1+2)+q^2(k_2+2)\right\}} 
\times SO(1)_1\rb
\times \cM_{k_1} \times \cM_{k_2} ~,  \nonumber \\
&&
\label{rewriting Tpq}
\end{eqnarray}
where $\cM_k$ denotes the $\cN=2$ minimal model with level $k$
($ c= \frac{3k}{k+2}$).
In the last line, we made use of the product formula of theta functions.
The $\cN=2$ worldsheet SUSY requires that
the part $\lb \cdots \rb$
becomes  the $\cN=2$ Liouville theory. 
The criticality condition \eqn{crit 2} gives us 
\begin{eqnarray}
&&  Q_{\phi}^2 = \frac{2(k_1+k_2+4)}{(k_1+2)(k_2+2)}~,
\end{eqnarray}
and hence if and only if $p=q$ holds, 
the correct factor $U(1)_{(k_1+2)(k_2+2)(k_1+k_2+4)}$ 
is obtained. 
In conclusion, only in the cases $p=q$, we have the $\cN=2$
worldsheet SUSY, leading to the $SU(3)$-holonomy.

This type string vacua were studied in \cite{Yamaguchi},  and
especially the simple cases of $k_1=k$, $k_2=0$ correspond to 
the $CY_3$ singularity of $A_{k+1}$-type \cite{GKP,ES1,ES2}
($D$, $E$-types are also described by the modular data).

We also make a comment on a geometric interpretation.
The Einstein homogeneous space $T^{p,q} = (SU(2)\times SU(2))/U(1)$ 
is defined by the $U(1)$-action
\begin{eqnarray}
e^{i\theta} ~\longmapsto ~ 
(e^{ip\th\frac{\sigma_3}{2}}, e^{iq\th\frac{\sigma_3}{2}} )~, 
\end{eqnarray}
with relatively prime integers $p$, $q$. 
It is well-known that only the case $p=q=1$ allows the spacetime SUSY
\cite{Romans}.
More precisely, $T^{p,q}$ becomes a Sasaki-Einstein space only for 
$p=q=1$. The cone over $T^{1,1}$ is the well-known conifold.
The parameters $p$, $q$ precisely correspond to those introduced in our 
discussions (at least for
the cases $k_1=k_2$). Thus the condition for the presence of spacetime SUSY 
in our construction is in agreement with geometrical considerations.

Such a correspondence was  partly suggested in \cite{ES2} and 
also discussed in \cite{PZT} in relation 
to the gauged WZW model of the GMM type \cite{GMM}.

Additionally we note that $T^{1,1} $ is isomorphic with the Stiefel manifold 
$V_2(\br^4)$ and have the $U(1)$-fibration
\begin{eqnarray}
T^{1,1}~\fiber{U(1)}~\CP_1\times \CP_1~, 
\end{eqnarray}
which gives the geometric interpretation of \eqn{rewriting Tpq}
for the case $p=q=1$.


~

\subsection{Coset Constructions of $d=3$ Superstring Vacua}

Since we are now assuming  $\dim G/H =6$,
the condition for the spacetime SUSY \eqn{cond non-compact G2}
reduces to the simple criterion given in \cite{SY2}:
The spacetime SUSY exists if and only if 
it holds that
\begin{eqnarray}
\cM &=& \frac{G_k\times SO(6)_1}{H} \cong \frac{G_k\times SU(3)_1}{H} \times
   \frac{SO(6)_1}{SU(3)_1}  \nonumber \\
& \cong & \frac{G_k\times SU(3)_1}{H} \times U(1)_{3/2} ~, 
\label{d=3 SUSY}
\end{eqnarray}
as a well-defined coset CFT.
Here we  used the equivalence
\begin{equation}
SO(6)_1/SU(3)_1 \cong U(1)_{3/2}~.
\end{equation}

We first focus on the examples when $G$ is a compact simple group.
The possible coset spaces are listed as follows;
\begin{eqnarray}
G/H&=&  SO(7)/SO(6)~,~SU(4)/(SU(3)\times U(1))~,~ G_2/SU(3)~, \nonumber \\
&& \hskip-10mm SO(5)/(SO(3)\times U(1))
~,~ SU(3)/(U(1)\times U(1))~.
\end{eqnarray}

We also pick up  the following example of a semi-simple $G$
\begin{eqnarray}
G/H=(SU(2)\times SU(2)\times SU(2))/SU(2)~.
\end{eqnarray}

~

\noindent
{\bf 1. $~ G/H=SO(7)/SO(6)~,~SU(4)/(SU(3)\times U(1))$}

The corresponding supercoset CFT's are  defined by
\begin{eqnarray}
&& \cM = \frac{SO(7)_k\times SO(6)_1}{SO(6)_{k+1}}~,~~~
\cM = \frac{SU(4)_k\times SO(6)_1}{SU(3)_{k+1}\times U(1)_{6(k+4)}}~. 
\end{eqnarray}
Clearly, the denominator groups are ``too big'' to allow the rearrangement \eqn{d=3 SUSY}.
We thus find no vacua with spacetime SUSY.

~

\noindent
{\bf 2. $~ G/H = G_2/SU(3)$}

In this case we find
\begin{eqnarray}
\cM &=& \frac{(G_2)_k\times SO(6)_1}{SU(3)_{k+1}} \nonumber \\
& \cong & \frac{(G_2)_k\times SU(3)_1}{SU(3)_{k+1}} \times U(1)_{3/2} ~. 
\end{eqnarray}
We thus obtain a supersymmetric vacuum corresponding to 
a non-compact $G_2$ holonomy manifold. This model has been studied in 
\cite{SY2}.

This coset space is isomorphic with $S^6$ and it is a typical example
of a nearly K\"{a}hler but non-K\"{a}hler manifold \cite{Gray}.
Our CFT result is in accordance with this fact.

~

\noindent
{\bf 3. ~ $G/H=SO(5)/(SO(3)\times U(1))$}

We have two possibilities of the embedding of $SO(3)\times U(1)$; 
one is to embed $SO(3)$ along the long root of $SO(5)$, 
and another is along its short root.
$U(1)$ has to be embedded in the direction orthogonal to $SO(3)$
in each case.

\begin{description}
 \item[(i) $SO(3)$ embedded along a long root :]

~

Suppose $SO(3)$ is embedded along a long root of $SO(5)$.
The coset space is topologically isomorphic with $S^7/U(1)\cong \CP^3$
and can be endowed with a (non-HSS) parabolic structure.\footnote
   {The most familiar coset realization of $\CP^3$ is of course 
    $SU(4)/(SU(3)\times U(1))$, which is a hermitian symmetric space
    (HSS).  The non-HSS K\"{a}hler coset $SO(5)/(SO(3)\times U(1))$ 
     gives an inequivalent parabolic decomposition 
    $$  \lieg = \lieh + \liem_+ + \liem_- ~, ~~~ 
     \lb \lieh,\liem_{\pm}\rb \subset \liem_{\pm}~,~~~ 
     \lb \liem_{\pm},\liem_{\pm} \rb \subset \liem_{\pm}~,
    $$
    with $\liem_+$, $\liem_-$ being non-abelian.
}
Hence $\cM$ is a (non-HSS) Kazama-Suzuki model. 
Based on the formulas \eqn{level formula} we obtain
\begin{eqnarray}
\cM&=&  \frac{SO(5)_k\times SO(6)_1}
{SO(3)_{\frac{k+1}{2}}\times U(1)_{k+3}} \nonumber \\
&\cong &  
\frac{SO(5)_k\times SO(6)_1}{SU(2)_{k+1}\times U(1)_{k+3}} \nonumber \\
&\cong &
\frac{SO(5)_k\times SU(3)_1}{SU(2)_{k+1}\times U(1)_{k+3}}
\times U(1)_{3/2}~.
\label{CP3 G2}
\end{eqnarray} 
We thus obtain superstring vacua of non-compact $G_2$-holonomy
manifolds. 

The coset space defined here can be equipped with an
Einstein metric that is $SO(5)$ invariant, but not compatible
with a definite complex structure. This is an example of a 
nearly K\"{a}hler manifold and gives rise to 
the well-known solution with the $G_2$-holonomy metric on its cone 
\cite{BS,GPP} (see also \cite{AW}). Our result seems consistent
with this fact.

\item[(ii) $SO(3)$ embedded along a short root : ]

~

If $SO(3)$ is embedded along a short root, the coset space is 
the Grassmannian $G_2(\br^5)$, which is an HSS. 
We find from \eqn{level formula}
\begin{eqnarray}
\cM &=&
\frac{SO(5)_k\times SO(6)_1}{SO(3)_{k+2} \times U(1)_{2(k+3)}} ~.
\end{eqnarray}
In this case we cannot rewrite it as in \eqn{d=3 SUSY}. 
The vacua have  no SUSY.

\end{description}

~

\noindent
{\bf 4.$ ~ G/H=SU(3)/(U(1)\times U(1))$ }

We must specify the charge lattice $\Gamma$ 
associated to the $U(1)\times U(1)$ action
to define the supercoset CFT $\cM$, although 
the corresponding coset space is 
always isomorphic with the flag manifold
$F_{1,2}(\bc^3)$ irrespective of the choice of $\Gamma$.
Let $Q$ be the root lattice of $SU(3)$. 
According to \eqn{level formula}, 
 $\Gamma$ must be a sublattice of $ \sqrt{k+3} \,Q$,
where $k$ is the level of $SU(3)$, in order to maintain 
the worldsheet SUSY.
On the other hand, the criterion for spacetime SUSY 
\eqn{d=3 SUSY} requires that the coset
\begin{eqnarray}
\frac{SU(3)_k\times SU(3)_1}{U(1)\times U(1)}
\end{eqnarray}
is well-defined, which implies the condition 
$\Gamma \subset \sqrt{k}Q \oplus Q$. 
Especially, we cannot adopt the simplest choice
$\Gamma = \sqrt{k+3} \,Q$ for a generic $k$.

The charge lattices $\Gamma$ which admit  both  
worldsheet and  spacetime SUSY
are constructed as follows: Let $\al_i$, $\beta_i$ ($i=1,2$)
be the simple roots corresponding to the current algebras $SU(3)_k$,
$SU(3)_1$ respectively, normalized as $\al_i^2=\beta_i^2=2$,
$(\al_1, \al_2)=(\beta_1,\beta_2)=-1$, $(\al_i,\beta_j)=0$.
We set
\begin{eqnarray}
\nu_1 &=& \sqrt{k}\al_1+ \beta_1+2\beta_2 ~, \nonumber\\  
\nu_2 &=& \sqrt{k}(\al_1+2\al_2) + 3\beta_1~,
\label{identification2}
\end{eqnarray}
and define the lattice
\begin{eqnarray}
\La = \bz \nu_1+\bz \nu_2~.
\label{lattice 1}
\end{eqnarray}
They satisfy 
\begin{eqnarray}
(\nu_1, \nu_1)=2(k+3)~,~~~(\nu_2,\nu_2)=6(k+3)~,~~~(\nu_1,\nu_2)=0~.
\end{eqnarray}
By definition $\La$ is a sublattice of  $ \sqrt{k}Q\oplus Q$,
and by means of the identification
\begin{eqnarray}
 \sqrt{k+3}\,\gamma_1 = \nu_1~,~~~ \sqrt{k+3}\,(\gamma_1+2\gamma_2) =\nu_2~,
\label{identification}
\end{eqnarray}
where $\gamma_1$, $\gamma_2$ denote the simple roots corresponding to
the total current $SU(3)_{k+3}$, we can also regard 
$\La$ as a sublattice of $ \sqrt{k+3}\,Q$.
More precise argument is given as follows:
Recall that the total Cartan current of $SU(3)_{k+3}$ is decomposed as 
\cite{KS,LVW}
\begin{eqnarray}
(t,H(z)) = (t,\hat{H}(z))
+\sum_{\al\in \Delta_+} \,\langle \al, t\rangle :\psi^{\al}\psi^{-\al}:~,
\end{eqnarray}  
where $\hat{H}$ is the bosonic current and $\psi^{\al}$ are the 
free fermions along the coset direction. 
$(\psi^{\al}(z)\psi^{\al'}(0) \sim \delta_{\al+\al',0}/z)$.
$\Delta_+$ denotes the set of positive roots.
On the other hand, the embedding $SU(3)_1\subset SO(6)_1 $
is given by defining the $SU(3)_1$ current $K(z)$ from the 6 free fermions
$\chi_i$, $\chi^*_i$ ($i=1,2,3$), 
($\chi^*_i(z)\chi_j(0)\sim \delta_{ij}/z$,
$\chi_i(z)\chi_j(0)\sim \chi^*_i(z)\chi^*_j(0)\sim 0$); 
\begin{eqnarray}
(E_{ii}-E_{jj}, K(z)) = :\chi^*_i\chi_i:-:\chi^*_j\chi_j:  ~, ~~~
(E_{ij}, K(z)) = \chi^*_i\chi_j ~~~(i\neq j)~,
\end{eqnarray}
where we set $(E_{ij})_{ab}=\delta_{ia}\delta_{jb}$.
Identifying the coset fermions $\psi^{\al}$ and $\chi_i$, $\chi^*_j$
by the relation
\begin{eqnarray}
&& \chi_1=\psi^{-\theta}~,~~~ \chi^*_1=\psi^{\theta} ~, \nonumber \\
&& \chi_2=\psi^{\al_2}~,~~~\chi^*_2=\psi^{-\al_2}~, \nonumber \\
&& \chi_3=\psi^{\al_1}~,~~~\chi^*_3=\psi^{-\al_1}~,
\end{eqnarray}
we find  that 
\begin{eqnarray}
(\la_3,H) &=& (\la_3,\hat{H}) + (:\chi^*_1\chi_1:+:\chi^*_2\chi_2:
-2:\chi^*_3\chi_3:) ~, \nonumber \\
(\la_8,H) &=& (\la_8,\hat{H}) + \sqrt{3}
(:\chi^*_1\chi_1:-:\chi^*_2\chi_2:) ~,
\end{eqnarray}
where $\la_3$, $\la_8$ are the Gell-Mann matrices corresponding to
$\al_1$, $ \frac{1}{\sqrt{3}}(\al_1+2\al_2)$;
\begin{equation}
\la_3 = \left(
\begin{array}{ccc}
 1& & \\
 & -1 & \\
 & & 0
\end{array}
\right)~,~~~
\la_8 = \frac{1}{\sqrt{3}}\left(
\begin{array}{ccc}
 1& & \\
 & 1 & \\
 & & -2
\end{array}
\right)~.
\label{Gell-Mann}
\end{equation}
These relations justify the identification 
\eqn{identification2}, \eqn{identification}.

We should also take account of the symmetry by the Weyl group $W$.
Therefore, the charge lattice $\Gamma$ gives SUSY, if (and only if)
it is a sublattice of $w\cdot \La$ with some Weyl reflection 
$w \in W$. Especially, in the simplest case 
$\Gamma = \La$, 
we obtain
\begin{eqnarray}
\cM&=& \frac{SU(3)_k\times SO(6)_1}{U(1)_{k+3}\times U(1)_{3(k+3)}} 
\nonumber\\
&\cong& \frac{SU(3)_k\times SU(3)_1}{U(1)_{k+3}\times U(1)_{3(k+3)}}
\times U(1)_{3/2}~.
\end{eqnarray}
One can directly check that the coset part in the last line 
is actually well-defined because of the relation
$SU(3)_1\sim U(1)_1\times U(1)_3 \sim U(1)_9\times U(1)_3$.
In conclusion, we have obtained an infinite family of
supersymmetric vacua of $G_2$-holonomy corresponding to
the charge lattice $\Gamma$ mentioned above.

We have a comment: The flag manifold $F_{1,2}(\bc^3)$ 
has the standard K\"{a}hler-Einstein metric, which is not
$SU(3)$-invariant. However, this space is known to have a second 
(non-K\"{a}hler) Einstein metric that is  $SU(3)$ invariant and 
compatible with a nearly K\"{a}hler structure \cite{Gray} (see also \cite{AW}).
There exists a well-known solution of $G_2$-holonomy metric on this cone 
\cite{BS,GPP}. Our CFT result again seems to be consistent with the classical
geometry. However, while the different choice of $\Gamma$ leads to the same homogeneous
space $F_{1,2}(\bc^3)$ classically, we obtain 
inequivalent string vacua depending on the choice of $\Gamma$ in our coset CFT 
construction.

~

We further consider examples of a non-simple $G$, motivated 
by the example of a nearly K\"{a}hler space $S^3\times S^3$.

~

\noindent
{\bf 5. $ ~ G/H=(SU(2)\times SU(2)\times SU(2))/SU(2)$}

We have three ways of $SU(2)$ embedding.  In all the three cases  
the coset space is topologically isomorphic with $S^3\times S^3$.

\begin{description}
 \item[(i) $SU(2)$ embedded only in an $SU(2)$-factor : ]

~

This case is very easy, since one of the $SU(2)$-factors
in the numerator is canceled.
The relevant supercoset reduces to
\begin{eqnarray}
\cM=SU(2)_{k_1}\times SU(2)_{k_2} \times SO(6)_1~.
\label{S3 S3 1}
\end{eqnarray} 
This obviously gives rise to supersymmetric vacua with 16 supercharges.

If we further make an $S^1$ compactification, this model will be 
converted by \eqn{interpolation} into the background
$AdS_3\times S^3 \times S^3 \times S^1$ discussed in \cite{EFGT},
which realizes the large $\cN=4$  superconformal symmetry
on the boundary of $AdS_3$.

~

\item[(ii) $SU(2)$ embedded in two $SU(2)$-factors :]

~

In this case the relevant supercoset becomes
\begin{eqnarray}
\cM=\frac{SU(2)_{k_1}\times SU(2)_{k_2} \times SO(3)_1}{SU(2)_{k_1+k_2+2}}
\times SU(2)_{k_3} \times SO(3)_1~.
\label{S3 S3 2}
\end{eqnarray} 
It leads to non-supersymmetric string vacua.

~

\item[(iii) $SU(2)$ embedded in all of the $SU(2)$-factors :]

~

This third case is the most interesting.
The relevant supercoset is defined as 
\begin{eqnarray}
\cM&=& \frac{SU(2)_{k_1}\times SU(2)_{k_2}\times SU(2)_{k_3}
\times SO(6)_1}{SU(2)_{k_1+k_2+k_3+4}} ~.
\label{S3 S3 3}
\end{eqnarray} 
In the same way as before
the existence of spacetime SUSY is confirmed by the equivalence
\begin{eqnarray}
\cM&\cong&\frac{SU(2)_{k_1}\times SU(2)_{k_2}\times SU(2)_{k_3}
\times SU(3)_1}{SU(2)_{k_1+k_2+k_3+4}} \times U(1)_{3/2} ~.
\label{coset relation 1}
\end{eqnarray}
Recall the conformal embedding $SU(2)_4 \subset SU(3)_1$ is
associated with the $D_4$-type modular invariant. Namely, 
we have the character relations
\begin{eqnarray}
&& \chi^{SU(3)_1}_{\msc{\bf basic}}(\tau) = \chi^{(4)}_0(\tau)
+\chi^{(4)}_4(\tau) \nonumber \\
&& \chi^{SU(3)_1}_{\msc{\bf fund}}(\tau) = 
\chi^{SU(3)_1}_{\overline{\msc{\bf fund}}}(\tau)= \chi^{(4)}_2(\tau)~, 
\end{eqnarray} 
where $\chi^{(k)}_{\ell}$ denotes the $SU(2)_k$ character of spin $\ell/2$.
Accordingly, the coset part in \eqn{coset relation 1} is really well-defined.

This type string vacua of $G_2$-holonomy are regarded as
natural generalizations of those given in \cite{ES3}. 
In fact, the special cases of $k_1=k$, $k_2=k_3=0$ reduces to
\begin{eqnarray}
\br_{\phi} \times \psi^{\phi} \times \frac{SU(2)_k\times SO(6)_1}{SU(2)_{k+4}}
&\cong& \br_{\phi} \times \psi^{\phi} \times
\frac{SU(2)_k\times SU(2)_2}{SU(2)_{k+2}}\times 
\frac{SU(2)_{k+2}\times SU(2)_2}{SU(2)_{k+4}} \nonumber \\
&\cong& \br_{\phi} \times \psi^{\phi} \times \cM^{\cN=1}_{k+2} 
\times \cM^{\cN=1}_{k+4}~,
\label{ES model}
\end{eqnarray}
where $\cM^{\cN=1}_m$ denotes the $m$-th $\cN=1$ minimal model
($c = \frac{3}{2}-\frac{12}{m(m+2)}$).
These are precisely the models constructed in \cite{ES3}.

We further make a comment: 
As mentioned in \cite{AW}, the trivial round sphere metric on 
$S^3\times S^3$ does not generate the $G_2$-holonomy on its cone.
While, the second Einstein metric based on the identification 
$S^3\times S^3 \cong (SU(2))^3/SU(2)$ leads to a $G_2$-holonomy 
on its cone and there exists a well-known $G_2$ metric \cite{BS,GPP}.
It seems plausible to relate the former case with the SCFT 
\eqn{S3 S3 1} and the latter with \eqn{S3 S3 3}, although
the precise interpretation of this relation is yet unclear.

\end{description}

~

To complete our classification of the $d=3$ string vacua,
let us also consider the diagonal coset \eqn{SU(2) diagonal}.
Although $\dim G/H\neq 6$,
we obtain the supersymmetric vacua of $G_2$-holonomy,
by restricting the levels of current algebras.
In fact, if we set $k_2=2$ (or $k_1=2$), 
the model again reduces to the one just considered \eqn{ES model}.  
It is easy to see that any other choices of levels do not lead to
supersymmetric vacua.

~


\subsection{Coset Constructions of $d=2$ Superstring Vacua}

The $d=2$ cases are the most involved  because 
we have three possibilities of supersymmetric string vacua
(except for the case of trivial flat space),
that is, the $Sp(2)$, $SU(4)$ and $Spin(7)$-holonomies,
each of which corresponds to a tri-Sasakian, Sasaki-Einstein, and 
weak $G_2$ base spaces, respectively.

The 7-dimensional Einstein homogeneous spaces are completely classified 
in \cite{CRW}. We first focus on the cases of simple $G$ listed as
\begin{eqnarray}
G/H&=& SO(8)/SO(7)~,~SO(7)/G_2~,~SU(4)/SU(3)~, \nonumber \\
&& ~SU(3)/U(1)~,~SO(5)/SO(3)~.
\end{eqnarray}

We further discuss a few cases with a non-simple  $G$;
$G/H=(SU(3)\times SU(2))/(SU(2)\times U(1))$,
$G/H= (SU(2)\times SU(2)\times SU(2))/(U(1)\times
U(1))$, which are also found in the list of \cite{CRW} 
(see also \cite{DNP}).
The other example is again \eqn{SU(2) diagonal}.
Although $\dim G/H \neq 7$ here, 
it will turn out that the $Spin(7)$-holonomy vacua are also obtained 
under a suitable restriction of levels, as in the $d=3$ case.


The SUSY condition 
\eqn{cond non-compact spin(7)} 
now reduces to a criterion \cite{SY2};
\begin{eqnarray}
\cM&=& \frac{G_k\times SO(7)_1}{H} \nonumber \\
 & \cong & \frac{G_k\times (G_2)_1}{H} \times \mbox{tri-critical Ising}~,
\label{d=2 SUSY}
\end{eqnarray}
by using the identification
\begin{equation}
SO(7)_1/(G_2)_1 \cong \mbox{tri-critial Ising}~,
\label{SV identity}
\end{equation}
which was first pointed out in \cite{SV}.

~

\noindent
{\bf 1. $ ~ G/H = SO(8)/SO(7)$ : }

We have no spacetime SUSY in this case, 
because the denominator group is too large to make the rearrangement
\eqn{d=2 SUSY} possible.

~


\noindent
{\bf 2. $ ~ G/H=SO(7)/G_2$ : }

This model gives the superstring vacua of $Spin(7)$-holonomy
studied in \cite{SY2}. 
The condition \eqn{d=2 SUSY} is easily checked as follows;
\begin{eqnarray}
\cM &=& \frac{SO(7)_k\times SO(7)_1}{(G_2)_{k+1}} \cong 
\frac{SO(7)_k\times (G_2)_1}{(G_2)_{k+1}} \times \frac{SO(7)_1}{(G_2)_{1}}
\nonumber \\
&\cong & \frac{SO(7)_k\times (G_2)_1}{(G_2)_{k+1}} \times 
\mbox{tri-critical Ising}~.  
\end{eqnarray}
The trivial case $k=0$ (i.e. $\cM= \mbox{tri-critical Ising}$) corresponds 
to the model discussed in \cite{ES3}.

~

\noindent
{\bf 3. $ ~ G/H=SU(4)/SU(3)$ : }

We can similarly prove that the condition \eqn{d=2 SUSY}
is satisfied here. However, based on \eqn{N=2 expression},
we can further show that the worldsheet SUSY is  enhanced to $\cN=2$;
\begin{eqnarray}
\br_{\phi} \times \psi^{\phi} \times \cM&=& \br_{\phi} \times \psi^{\phi} 
\times \frac{SU(4)_k\times SO(7)_1}{SU(3)_{k+1}} \nonumber \\
&\cong& \left\lb \br_{\phi} \times \psi^{\phi} \times 
   U(1)_{6(k+4)} \times SO(1)_1 \right\rb \times 
  \frac{SU(4)_k\times SO(6)_1}{SU(3)_{k+1}\times U(1)_{6(k+4)}} ~.
\end{eqnarray}
The part $\lb \cdots \rb$ is interpreted as the $\cN=2$ Liouville theory. 
The criticality condition \eqn{crit 2} gives  
$  Q_{\phi}^2 = 12/(k+4)$, 
and hence $U(1)_{6(k+4)}$ describes precisely the compact boson 
of $\cN=2$ Liouville theory.
The remaining coset CFT is the Kazama-Suzuki model 
for $\CP_3$. Therefore, these string vacua correspond to 
non-compact $CY_4$ manifolds.

~

\noindent
{\bf 4.$ ~ G/H=SO(5)/SO(3)$ :}


This example is quite amazing. We find that all of the three possible holonomies
$Sp(2)$, $SU(4)$, and $Spin(7)$ are realized. 
\begin{description}
 \item[(i) $SO(3)$ embedded along a long root :] 

~

Suppose  $SO(3)$ is embedded along a long root of $SO(5)$,
in other words, embedded in one of 
the $SO(3)$'s of the 
$SO(3)\times SO(3) (\cong SO(4))$ subgroup of $SO(5)$.
This coset space is isomorphic with $S^7$, and as is obvious 
by construction,  we have a remaining $SO(3) (\simeq SU(2))$
symmetry. This space is an elementary example of the tri-Sasakian 
manifold and can be regarded as an $SU(2)$-bundle over a Wolf space:
\begin{eqnarray}
\cM= SO(5)/SO(3) \cong S^7~
\stackrel{SU(2)}{\longrightarrow}~ \cM/SU(2) \cong S^4~.
\end{eqnarray}
This  is nothing but  the familiar (quaternionic) Hopf fibration.

In terms of the coset CFT, we obtain  
\begin{eqnarray}
\br_{\phi} \times \psi^{\phi} \times \cM&=& 
\br_{\phi}\times \psi^{\phi}\times 
\frac{SO(5)_k\times SO(7)_1}{SO(3)_{\frac{k+1}{2}}} \nonumber \\
&\cong& \br_{\phi}\times \psi^{\phi}\times 
\frac{SO(5)_k\times SO(7)_1}{SU(2)_{k+1}}  \nonumber \\
&\cong&
\left\lb \br_{\phi}\times \psi^{\phi}\times SU(2)_{k+1} \times SO(3)_1 
\right\rb
\times \frac{SO(5)_k\times SO(4)_1}{SU(2)_{k+1}\times SU(2)_{k+1}}~.
\nonumber \\
&& \hspace{10cm}
\label{3S SO}
\end{eqnarray} 
The criticality condition \eqn{crit 2} yields 
$Q_{\phi}^2={8}/(k+3)$. Since the supercoset part of the last line 
is associated to a Wolf space, the worldsheet SUSY should be enhanced 
to $\cN=4$, as we already discussed.
In this way, we have obtained the superstring vacua corresponding to 
the $Sp(2)$-holonomy. The SUSY cancellation reduces to the identity
\eqn{cb cy3}.

~

\item[(ii) $SO(3)$ embedded along a short root :]

~

Suppose $SO(3)$ is embedded along a short root, in other words,
embedded diagonally into the $SO(3)\times SO(3)$ subgroup.
This is found to be the ``canonical embedding'' so that the vector 
representation of $SO(5)$ is decomposed as ${\bf 5}\,\rightarrow 
{\bf 3} + {\bf 1} +{\bf 1}$. Hence the coset space is 
isomorphic with the Stiefel manifold $V_2(\br^5)$, 
which has the canonical $U(1)(\cong SO(2))$-fibration over the 
Grassmannian $G_2(\br^5)$ (an example of HSS);
\begin{eqnarray}
\cM=SO(5)/SO(3)\cong V_2(\br^5)~\stackrel{U(1)}{\longrightarrow}~
\cM/U(1) 
\cong G_2(\br^5) ~.
\end{eqnarray}
This is a typical example of the Sasaki-Einstein homogeneous space. 
In terms of the coset CFT we thus find that
\begin{eqnarray}
&&\br_{\phi} \times \psi^{\phi} \times \cM
= \br_{\phi}\times \psi^{\phi}\times 
\frac{SO(5)_k\times SO(7)_1}{SO(3)_{k+2}} \nonumber \\
&&\hskip10mm \cong
\left\lb \br_{\phi}\times \psi^{\phi}\times U(1)_{2(k+3)} 
\times SO(1)_1 \right\rb
\times \frac{SO(5)_k\times SO(6)_1}{SO(3)_{k+2}\times U(1)_{2(k+3)}}.
\end{eqnarray}
We here obtain $Q_{\phi}^2={9}/(k+3)$, resulting in  
$ 3^2 \cdot 2/Q_{\phi}^2= 2(k+3)$. Therefore, $U(1)_{2(k+3)}$-factor
exactly reproduces the $\cN=2$ Liouville theory. The remaining coset CFT 
is the Kazama-Suzuki model associated to $G_2(\br^5)$. 
In this way we find that the total system has $\cN=2$ worldsheet 
SUSY and $SU(4)$-holonomy.

~

\item[(iii) $SO(3)$ embedded as a maximal subgroup of $SO(5)$:]

~

The third case is the most non-trivial.
Consider the embedding of $SO(3)$ along $2\al_1+3\al_2$,
where $\al_1$, $\al_2$ are the long and short roots of $SO(5)$
($\al_1^2=2$, $\al_2^2=1$, $\al_1\cdot\al_2=-1$).
The simple root of $SO(3)$ is defined as the projection of 
the highest root of $SO(5)$ and thus identified as 
$ \th'\equiv \frac{1}{5}(2\al_1+3\al_2)$. 
In this embedding the adjoint representation of $SO(5)$ is decomposed as 
${\bf 10} ~\rightarrow~ {\bf 3}+{\bf 7}$, which means  it is a maximal
embedding (see \cite{CRW} for the detail).  
Since we have $(\th',\th')=1/5$, the relevant supercoset should be
\begin{eqnarray}
\cM= \frac{SO(5)_k\times SO(7)_1}{SO(3)_{5k+14}}~.
\label{SO(5)-SO(3) max}
\end{eqnarray}
Since we have no remaining symmetry of $SU(2)$ or $U(1)$ in this coset, 
it is obvious that the worldsheet SUSY cannot be enhanced.
We thus obtain at most 2 supercharges in spacetime corresponding to
the $Spin(7)$-holonomy.
The criterion for the spacetime SUSY \eqn{d=2 SUSY} is now 
expressed as 
\begin{eqnarray}
\cM &\cong& \frac{SO(5)_k\times (G_2)_1}{SO(3)_{5k+14}} \times 
\frac{SO(7)_1}{(G_2)_1} \nonumber \\
&\cong & \frac{SO(5)_k\times (G_2)_1}{SU(2)_{10k+28}} \times 
\mbox{tri-critical Ising}~,
\label{SO(3) weak G2}
\end{eqnarray}  
and we require that the coset CFT in the last line should
be well-defined. Especially, we must ask whether we can consistently 
define $(G_2)_1/SU(2)_{28}$. We first note that $(G_2)_1$ and
$SU(2)_{28}$ have the equal central charge $c=14/5$. So,
$(G_2)_1/SU(2)_{28}$ would be a topological coset CFT, if it is well-defined.
It is actually known that the conformal embedding 
$SU(2)_{28}\subset (G_2)_1$ exists (see for example,
\cite{CFTtext})\footnote
              {The explicit embedding of $SU(2) \subset G_2$ here
      is given as follows: The simple root of $SU(2)$ is identified with  
       $\th' = \frac{1}{14}(\al_1+6\al_2)$, 
       where $\al_1$, $\al_2$ denote the long and short
      roots of $G_2$ ($\al_1^2=2$, $\al_2^2=2/3$, $\al_1\cdot\al_2=-1$).
       By this embedding, the adjoint representation of $G_2$ is
      decomposed as $\bf 14 ~ \rightarrow~  3+ 11$. 
       Hence, this is also the maximal embedding. 
       The square length of $\th'$ is equal $1/14$, 
       which is compatible with the existence of 
       coset CFT $(G_2)_1/SU(2)_{28}$.}.
More precise relation is as follows: $SU(2)_{28}$ is known
to have the $E_8$-type modular invariant \cite{CIZ},
in which the partition function is given as
\begin{eqnarray}
&& Z=\left|\chi^{(28)}_0+\chi^{(28)}_{10}+\chi^{(28)}_{18}+
\chi^{(28)}_{28}\right|^2
+\left|\chi^{(28)}_6+\chi^{(28)}_{12}
+\chi^{(28)}_{16}+\chi^{(28)}_{22}\right|^2~.
\end{eqnarray}
This partition function is in fact equivalent to 
the diagonal modular invariant of $(G_2)_1$ due to
the character relations;
\begin{eqnarray}
\chi^{(G_2)_1}_{\msc{\bf basic}}(\tau) &=& 
\left(\chi^{(28)}_0+\chi^{(28)}_{10}+
\chi^{(28)}_{18}+\chi^{(28)}_{28}\right) (\tau)~,
\nonumber \\
\chi^{(G_2)_1}_{\msc{\bf fund}}(\tau) &=& \left(
\chi^{(28)}_6+\chi^{(28)}_{12}+\chi^{(28)}_{16}+\chi^{(28)}_{22}\right)(\tau)~.
\end{eqnarray}
Accordingly, the rewriting \eqn{SO(3) weak G2} 
is actually well-defined. 
We have achieved the string vacua of $Spin(7)$-holonomy.

\end{description}

~

\noindent
{\bf 5. $ ~G/H=SU(3)/U(1)$ : }

To define the supercoset CFT, we have to specify
the $U(1)$-embedding, which is characterized by the 1-dimensional 
charge lattice $\Gamma \subset \sqrt{k+3}\,Q$, where $Q = \bz \al_1+\bz \al_2$
is the root lattice of $SU(3)$. ($(\al_i, \al_i)=2$, $(\al_1,\al_2)=-1$.)

The spacetime SUSY requires that the following rewriting should be possible; 
    \footnote{This is a slightly stronger condition 
     than \eqn{d=2 SUSY} and leads to twice as many supercharges 
     as compared to the $Spin(7)$-holonomy case
     (i.e. the same number of supercharges as the $SU(4)$ and $G_2$-holonomy).
     However, it is easy to show that \eqn{d=2 SUSY} inevitably 
     reduces to \eqn{rewrite 1} in this case.}
\begin{eqnarray}
\cM & =& \frac{SU(3)_k\times SO(7)_1}{U(1)} \nonumber\\
& \cong & \frac{SU(3)_k\times SU(3)_1}{U(1)} \times SO(1)_1
 \times U(1)_{3/2} ~.
\label{rewrite 1}
\end{eqnarray}
As we discussed in the $d=3$ analysis, this rewriting is possible 
if and only if $\Gamma$ is 
generated by an element $\mu_1$ of the lattice $w\cdot \La$,
where $\La$ is defined in \eqn{lattice 1} and $w$ is a Weyl reflection.
However, we can show that $\sqrt{k+3}\,Q = W\cdot \La$. 
So, we have the spacetime SUSY for an arbitrary choice
of $\mu_1$.

On the other hand, the $\cN=4$ enhancement of worldsheet SUSY
occurs in the special cases $\mu_1= m(w\cdot \nu_2) 
(\equiv m(w\cdot (\al_1+2\al_2)))$, where $m$ is an arbitrary integer
and $w$ is a Weyl reflection.
In fact, (only) in that case we can find an $SU(2)$ symmetry along 
the transverse direction $\mu_2=w\cdot \nu_1(\equiv w\cdot \al_1)$,
resulting in the equivalence
\begin{eqnarray}
\br_{\phi}\times \psi^{\phi}\times \cM &\cong&
\left\lb \br_{\phi}\times \psi^{\phi}\times  SU(2)_{k+1} \times
SO(3)_1\right\rb \times \frac{SU(3)_k\times SO(4)_1}
{U(1)_{3(k+3)} \times SU(2)_{k+1}}~.
\label{N11 N=4}
\end{eqnarray} 
The coset part is associated to $\CP_2$, which possesses  the structures 
of both the HSS and Wolf space. Hence the worldsheet SUSY
is enhanced to $\cN=4$. 

As a consistency check,
we can also check the $\cN=2$ structure on the worldsheet.
The following rewriting is also possible;
\begin{eqnarray}
\br_{\phi}\times \psi^{\phi}\times \cM &\cong&
\left\lb \br_{\phi}\times \psi^{\phi}\times  U(1)_{k+3} \times
SO(1)_1\right\rb \times \frac{SU(3)_k\times SO(6)_1}
{U(1)_{3(k+3)} \times U(1)_{k+3}}~.
\label{N11 N=2}
\end{eqnarray} 
The part $\lb \cdots \rb$ describes the $\cN=2$ Liouville
since $Q_{\phi}^2 =8/(k+3)$, and 
the coset part is the Kazama-Suzuki model considered in 
the $d=3$ analysis.

It is also  obvious that,
if we cannot write $\mu_1$ as the form 
$\mu_1=m(w\cdot \nu_2)$, the worldsheet SUSY cannot be enhanced
for generic values of $k$.

In summary, we have shown that  
\begin{description}
 \item[(i)] If $\Gamma$ has the form $\Gamma = \bz (m w\cdot\nu_2)$,
with some Weyl reflection $w$ and an integer $m$,
we have the $\cN=4$ worldsheet SUSY, and 
the string vacua corresponds to an $Sp(2)$-holonomy.
 \item[(ii)] If $\Gamma$ does not, we have the supersymmetric vacua
with 4 supercharges, but the worldsheet SUSY is at most $\cN=1$.
In this case, the coset CFT $\cM$ has a remaining $U(1)$-symmetry and  
seems to correspond to a compactification on a space of the form
$S^1\times \mbox{$G_2$-manifold}$.  
However, the $U(1)$-charge is not independent 
of the quantum numbers in the remaining sector and this is
some kind of an orbifold space.   
\end{description}

Let us  make a few comments: 
The coset space $SU(3)/U(1)$ is known under the name 
Aloff-Wallach space \cite{Aloff} and written as  $N(m,\ell)$ \footnote
      { In many literature
         it is also denoted as $N^{pqr}$ \cite{CR}, where 
       $ \dsp
N^{pqr} = \frac{SU(3)\times U(1)}{U(1)\times U(1)}
$ with the integer parameters $p$, $q$, $r$
characterizing the remaining $U(1)$ symmetry as 
$$
Z=p\frac{i\sqrt{3}}{2}\la_8+q\frac{i}{2}\la_3 +riY~.
$$
($\la_3$, $\la_8$ are the Gell-Mann matrices \eqn{Gell-Mann}
and $Y$ is the generator of
$U(1)$ in the numerator.) In particular, $N(1,1)$ is equal to $N^{010}$.
} with the $U(1)$-action 
\begin{equation}
e^{i\theta}~\longmapsto~
\left(
\begin{array}{ccc}
 e^{im\theta}& 0  & 0 \\
 0& e^{i\ell\theta} & 0 \\
 0&0 & e^{-i(m+\ell)\theta}
\end{array}
\right)
\label{AW space}
\end{equation}
where $m$, $\ell$ are relatively prime integers.
They are not diffeomorphic for different parameters $m$, $\ell$
(unless we have a Weyl reflection connecting them).

It is known that the spaces $N(m,\ell)$ can be endowed  with  
two types of Einstein metrics \cite{CR,PP}, 
denoted as $N(m,\ell)_{\msc{I}}$ and $N(m,\ell)_{\msc{II}}$ 
(the ``squashed'' one) in \cite{DNP}. 
The generic cases of $(m,\ell) \neq (1,1)$ become 
weak $G_2$ manifolds irrespective of the choice of Einstein metrics.
On the other hand, $N(1,1)_{\msc{I}}$ is known to be tri-Sasakian (and
also Sasaki-Einstein) \cite{CR,BG},  
while  $N(1,1)_{\msc{II}}$ has the  weak $G_2$ holonomy \cite{PP,BG}.
The choice of $m$, $\ell$ is in one-to-one correspondence with 
the charge lattice $\Gamma$ introduced above.
The first case (i) corresponds to $N(1,1)$ and leads to the $\cN=4$
worldsheet SUSY, while the second case (ii) corresponds to the cases 
$(m,\ell)\neq (1,1)$ and at most the $\cN=1$ worldsheet SUSY is allowed.
In this sense our algebraic construction agrees with 
the geometrical analysis. The amount of worldsheet SUSY is 
exactly as expected. Among other things, $N(1,1)_{\msc{I}}$
has the $SU(2)$-fibration characteristic of the tri-Sasakian homogeneous space;
\begin{eqnarray}
N(1,1)_{\msc{I}}~\fiber{SU(2)}~\CP_2~,
\end{eqnarray}  
which corresponds to \eqn{N11 N=4}.
It also has the $U(1)$-fibration for  the Sasaki-Einstein space;
\begin{eqnarray}
N(1,1)_{\msc{I}}~\fiber{U(1)}~F_{1,2}(\bc^3)~,
\end{eqnarray} 
where we should regard the flag manifold $F_{1,2}(\bc^3)$ as a 
K\"{a}hler-Einstein space. This of course corresponds to 
\eqn{N11 N=2}.

We will later discuss the CFT 
interpretation of the squashed geometry of $N(1,1)_{\msc{II}}$.

~

We further analyse a few examples with a non-simple group $G$.

~

\noindent
{\bf 6. $ ~ G/H=(SU(3)\times SU(2))/(SU(2)\times U(1))$}

We have  various possibilities of embedding of $SU(2)\times U(1)$ as listed 
in \cite{CRW}.

\begin{description}
 \item[(i) $SU(2)$ embedded as the isospin subgroup :]

~

The relevant coset SCFT should have the form
\begin{eqnarray}
\cM&=& \frac{SU(3)_{k_1}\times SU(2)_{k_2} \times SO(7)_1}
{SU(2)_{k_1+1}\times U(1)}~.
\label{coset Mpq 1}
\end{eqnarray}
To define the model completely, we still have to fix the $U(1)$ 
embedding. 
Choosing the isospin $SU(2)$ subgroup along the simple root $\al_1$
(simple roots of $SU(3)$ : $\al_1$, $\al_2$ with
 $\al_1^2=\al_2^2=2$, $\al_1\cdot \al_2=-1$, as usual),
let us introduce the following charge lattice for $U(1)$-action;
\begin{eqnarray}
\Gamma &=& \bz\left(q\sqrt{k_1+3}(\al_1+2\al_2)-p'\sqrt{k_2+2}
  \beta\right)~,
\label{gamma p q}
\end{eqnarray}
where $\beta$ is the simple root of the $SU(2)$ factor ($\beta^2=2$).
The $U(1)$ action
generated by $\Gamma$ obviously commutes with 
$SU(2)_{\msc{isospin}}$ for arbitrary $q$, $p'$.
The theta function associated to this charge lattice yields 
the $U(1)$ factor $U(1)_{3(k_1+3)q^2+(k_2+2){p'}^2}$ in  
\eqn{coset Mpq 1}, and we obtain
\begin{eqnarray}
\cM &\cong& \frac{SU(3)_{k_1}\times SO(4)_1}
{SU(2)_{k_1+1}\times U(1)_{3(k_1+3)}} \times 
\frac{SU(2)_{k_2}\times SO(2)_1}{U(1)_{k_2+2}} \nonumber \\
&& ~~~~~
\times SO(1)_1 \times U(1)_{3(k_1+3)(k_2+2)
\left\{3q^2(k_1+3)+{p'}^2(k_2+2)\right\}}~,
\label{rewrite Mpq}
\end{eqnarray}
using the relation
\begin{eqnarray}
\frac{U(1)_{3(k_1+3)}\times U(1)_{k_2+2}}{U(1)_{3(k_1+3)q^2+(k_2+2){p'}^2}}
\cong U(1)_{3(k_1+3)(k_2+2)\left\{3q^2(k_1+3)+{p'}^2(k_2+2)\right\}}~,
\end{eqnarray}
of the product formula of theta functions.
The first coset part in \eqn{rewrite Mpq} is the Kazama-Suzuki model 
for $\CP_2$ and the second coset is the $\cN=2$ minimal model of level 
$k_2$. On the other hand, the criticality condition \eqn{crit 2} leads to
\begin{eqnarray}
Q_{\phi}^2 = \frac{2(k_1+3k_2+9)}{(k_1+3)(k_2+2)}~.
\end{eqnarray}
We thus need  the factor $U(1)_{(k_1+3)(k_2+2)(k_1+3k_2+9)}$ in 
\eqn{rewrite Mpq}
to get the $\cN=2$ Liouville sector. 
Therefore, the worldsheet SUSY is  enhanced  to $\cN=2$,
if and only if $p'=3q$ holds, which yields a string vacuum of 
$SU(4)$-holonomy. 

In the case of $p'\not =3q$, we have at most $\cN=1$
worldsheet SUSY. Then, the spacetime SUSY requires that the 
following rewriting should be possible
(as in the analysis of $SU(3)/U(1)$);
\begin{eqnarray}
\cM &\cong& \frac{SU(3)_{k_1}\times SU(2)_{k_2}\times SU(3)_1}
{SU(2)_{k_1+1}\times U(1)_{3q^2(k_1+3)+{p'}^2(k_2+9)}} \times SO(1)_1
\times U(1)_{3/2}~.
\end{eqnarray}
For generic values of $k_1$, $k_2$, this is possible only for $p'=3q$,
which goes back to the $\cN=2$ case already discussed.
In this way, we conclude that this type string vacua are supersymmetric 
if and only if $p'=3q$ holds, and in those cases we obtain the $SU(4)$-holonomy.

We have a comment in connection to known results in Kaluza-Klein SUGRA:
This case corresponds to the homogeneous space with 
$SU(3)\times SU(2)\times U(1)$ isometry first studied in \cite{Witten-G};
\begin{eqnarray}
M^{pqr} &=& \frac{SU(3)\times SU(2) \times U(1)}
{SU(2)\times U(1) \times U(1)}~,
\end{eqnarray}
where $SU(2)$ is embedded as the isospin subgroup of $SU(3)$ and
the integer parameters $p$, $q$, $r$ characterize the remaining 
$U(1)$-symmetry as
\begin{eqnarray}
Z=p\frac{i\sqrt{3}}{2}\la_8+q\frac{i}{2}\sigma_3 +riY~,
\end{eqnarray}
where $\la_8$ is the Gell-Mann matrix (the generator transverse to
the $SU(2)_{\msc{isospin}}$) \eqn{Gell-Mann}, 
$\sigma_3$ is the Pauli matrix and
$Y$ is the generator of the $U(1)$-factor in the numerator.
Our coset CFT $\cM$ with the charge lattice $\Gamma$
is naturally associated to the space $M^{pq0}$ 
under the identification $p'=3p$
(at least for the cases
$k_1+3=k_2+2$). It is known \cite{CDF} that every coset space $M^{pqr}$
can be equipped with Einstein metrics and allows the spacetime SUSY
only in the case $p=q$. $M^{110}$ is a regular Sasaki-Einstein space and 
$M^{ppr}$ is regarded as an orbifold of it.
The $U(1)$-fibration for the Sasaki-Einstein space $M^{110}$
is written as 
\begin{eqnarray}
M^{110}~\fiber{U(1)}~\CP_2\times \CP_1~,
\end{eqnarray} 
which corresponds  to \eqn{rewrite Mpq} for $q=1,p'=3$.   
These aspects fit nicely with our algebraic construction.

~

\item[(ii) $SU(2)$ embedded in the explicit $SU(2)$ factor : ]

~

Obviously, this reduces to the case of $SU(3)/U(1)$ we studied
previously.

~

\item[(iii) $SU(2)$ embedded in both of $SU(3)$
and the explicit $SU(2)$ factor : ]

~

In this case the only possibility is the diagonal embedding 
$SU(2)\subset SU(2)_{\msc{isospin}}\times SU(2) \subset SU(3)\times SU(2)$.
The $U(1)$-factor must necessarily be embedded along the 
$\la_8 \propto \al_1+2\al_2$ direction.
The corresponding coset SCFT is defined as 
\begin{eqnarray}
\cM&=& \frac{SU(3)_{k_1} \times SU(2)_{k_2} \times SO(7)_1}
{SU(2)_{k_1+k_2+3}\times U(1)_{3(k_1+3)}} ~.
\label{squashed N11}
\end{eqnarray}
We can rewrite it as 
\begin{eqnarray}
\cM &\cong& \frac{SU(3)_{k_1}\times SU(2)_{k_2} \times (G_2)_1}
{SU(2)_{k_1+k_2+3} \times U(1)_{3(k_1+3)}} \times \mbox{tri-critial Ising}~.
\end{eqnarray}
(Recall that $(G_2)_1 \sim SU(2)_3 \times U(1)_1 \sim SU(2)_3 \times U(1)_9$.)
Therefore, we have obtained a string vacuum with $Spin(7)$-holonomy.

The coset space considered here  is topologically 
isomorphic with $N(1,1)$, and this string vacuum is supposed to 
correspond to the squashed geometry of $N(1,1)$ mentioned before. 
Under the limit $k_2\, \rightarrow \,  \infty$, the $SU(2)$-factors  
in \eqn{squashed N11} are canceled out, and we recover the 
$N(1,1)$ coset SCFT $\dsp \cM
=  \frac{SU(3)_k \times SO(7)_1}{U(1)_{3(k+3)}}$.

~

\item[(iv) $SU(2)$ embedded as a maximal subgroup of $SU(3)$ : ]

~

In this case $U(1)$ has to be embedded only in the 
$SU(2)$ factor. As in \eqn{SU(3)-SU(2) max}, the relevant supercoset CFT 
is written as 
\begin{eqnarray}
\cM&=& \frac{SU(3)_{k_1}\times SU(2)_{k_2}\times SO(7)_1}
{SU(2)_{4k_1+10} \times U(1)_{k_2+2}} \\
\frac{}{} &\cong & 
\frac{SU(3)_{k_1}\times SO(5)_1}{SU(2)_{4k_1+10}} \times 
\frac{SU(2)_{k_2}\times SO(2)_1}{U(1)_{k_2+2}} ~. \nonumber 
\end{eqnarray}
The second line corresponds to the fact that 
this coset space is topologically isomorphic with $S^5\times S^2$
(see \cite{CRW}), and no remaining $U(1)$ symmetry exists.
As is easily shown, we have no spacetime SUSY in this case.
It is again consistent with the known results of Kaluza-Klein SUGRA 
\cite{CRW}.

\end{description}

~

\noindent
{\bf 7. $~ G/H=(SU(2)\times SU(2)\times SU(2))/(U(1)\times U(1))$ : }

This model may regarded as a natural generalization of the $d=4$ vacuum
$\dsp (SU(2)\times SU(2))/U(1)$ and also the CHS $\sigma$-model. 
The supercoset has the form
\begin{eqnarray}
\cM=\frac{SU(2)_{k_1}\times SU(2)_{k_2}\times SU(2)_{k_3}\times SO(7)_1}
{U(1)\times U(1)}~.
\end{eqnarray}
Similarly to the $d=4$ case, 
we try to rewrite as 
\begin{eqnarray}
\br_{\phi}\times \psi^{\phi} \times \cM &\sim &
\left\lb \br_{\phi} \times \psi^{\phi} \times U(1)\times SO(1)_1
\right\rb \times 
\cM_{k_1}\times \cM_{k_2}\times \cM_{k_3}~,
\label{rewrite Qpqr} 
\end{eqnarray}  
where $\dsp \cM_k \equiv \frac{SU(2)_k\times SO(2)_1}{U(1)_{k+2}} $
denotes the $\cN=2$ minimal model again.
Since  we obtain 
\begin{eqnarray}
  Q_{\phi}^2= \frac{2(N_2N_3+N_3N_1+N_1N_2)}{N_1N_2N_3}~,~~~ 
(N_i\equiv k_i+2)~,
\end{eqnarray}
from the criticality condition \eqn{crit 2},
the criterion for the part $\lb \cdots \rb$ to become 
the $\cN=2 $ Liouville 
is whether we can factorize $U(1)_{N_1N_2N_3(N_2N_3+N_3N_1+N_1N_2)}$.
Namely, we want to derive a relation 
\begin{eqnarray}
\frac{U(1)_{N_1}\times U(1)_{N_2} \times U(1)_{N_3}}{U(1)\times U(1)}
\cong U(1)_{N_1N_2N_3(N_2N_3+N_3N_1+N_1N_2)}~,
\label{relation 1}
\end{eqnarray}
by a suitable choice of the charge lattice $\Gamma$ of $U(1)\times U(1)$.
To describe it explicitly,  let $\al_i$ ($i=1,2,3$) be the simple roots of 
each $SU(2)$ factors, normalized as $(\al_i,\al_j)=2\delta_{ij}$.
The two dimensional lattice $\Gamma$ must be defined as a sublattice
of $\bz \sqrt{N_1}\al_1+\bz \sqrt{N_2}\al_2+\bz \sqrt{N_2}\al_2$.
Introducing integer parameters $p$, $q$, $r$,
we set 
\begin{eqnarray}
\nu_1 &=& q\sqrt{N_1}\al_1-p\sqrt{N_2}\al_2~, \nonumber\\
\nu_2 &=& prN_2N_3\sqrt{N_1}\al_1+qrN_3N_1\sqrt{N_2}\al_2
-N_3(p^2 N_2+q^2 N_1)\sqrt{N_3}\al_3~, \nonumber\\
\nu_3 &=& pN_2N_3\sqrt{N_1}\al_1+qN_3N_1\sqrt{N_2}\al_2
+rN_1N_2\sqrt{N_3}\al_3~.
\end{eqnarray}
Then we find that they are orthogonal to each other, 
and also,
\begin{eqnarray}
(\nu_3, \nu_3)= 2N_1N_2N_3(p^2N_2N_3+q^2N_3N_1+r^2N_1N_2)~.
\end{eqnarray}
If we choose $\Gamma$ as (a sublattice of) 
$\bz \nu_1 + \bz \nu_2$, we obtain the theta function identity 
such as
\begin{eqnarray}
\Th{*}{N_1}(\tau)\Th{*}{N_2}(\tau)\Th{*}{N_3}(\tau)
= \sum \Theta^{(\Gamma)}_{*}(\tau) 
\Th{*}{N_1N_2N_3(p^2N_2N_3+q^2N_3N_1+r^2N_1N_2)}~.
\end{eqnarray}
Accordingly, if and only if $p=q=r$ holds, the wanted relation
\eqn{relation 1} is obtained and hence the worldsheet SUSY enhances 
to $\cN=2$. We find the string vacuum with $SU(4)$-holonomy
in this case. It is also not difficult to see that the spacetime SUSY
cannot exist in other cases as in the previous analysis.
The special cases of $k_1=k$, $k_2=k_3=0$ correspond to the $CY_4$
with the $A_{k+1}$-type singularity studied in \cite{GKP,ES1,ES2}.

We also make a comment in connection to the Kaluza-Klein SUGRA:
Consider the Einstein homogeneous space \cite{DFN} 
\begin{eqnarray}
Q(p,q,r) = \frac{SU(2)\times SU(2) \times SU(2)}{U(1)\times U(1)}~,
\end{eqnarray}
where $p$, $q$, $r$ parameterize the remaining $U(1)$-symmetry 
as before. 
It is known that we have the spacetime SUSY
only in the cases $p=q=r$ and the $Q(p,q,r)$ space becomes
a Sasaki-Einstein space  in this case. 
For the cases of $N_1=N_2=N_3$, the above choice of charge lattice 
$\Gamma$ precisely reproduces the coset space $Q(p,q,r)$ (the vector
$\nu_3$ describes the remaining $U(1)$ symmetry), and 
thus the SUSY condition coincides precisely.
Moreover, the $U(1)$-fibration
\begin{eqnarray}
Q(p,q,r) ~\fiber{U(1)}~\CP_1\times \CP_1 \times \CP_1
\end{eqnarray}
is naturally related with \eqn{rewrite Qpqr}. 
Our CFT analysis again is  consistent with 
that of Kaluza-Klein SUGRA.

~

\noindent
{\bf  8. $ ~ G/H=(SU(2)\times SU(2))/SU(2)$ : }

Finally we present an example with $\dim G/H\neq 7$.
Consider again the $\cN=1$ diagonal coset \eqn{SU(2) diagonal}
and set $k_1=k$, $k_2=1$;
\begin{eqnarray}
\cM= \frac{SU(2)_k\times SU(2)_1\times SO(3)_1}{SU(2)_{k+3}} ~.
\end{eqnarray}
Since $\dim G/H=3$ rather than $\dim G/H=7$,  
we use a different relation to present the spacetime SUSY;
\begin{eqnarray}
\frac{SU(2)_1\times SU(2)_2}{SU(2)_3} \cong \mbox{tri-critical Ising}~.
\end{eqnarray}
We then obtain
\begin{eqnarray}
\cM &\cong & \frac{SU(2)_k\times SU(2)_3}{SU(2)_{k+3}} \times 
\frac{SU(2)_1\times SU(2)_2}{SU(2)_3} \nonumber \\
&\cong &
\frac{SU(2)_k\times SU(2)_3}{SU(2)_{k+3}} \times
\mbox{tri-critial Ising}~,
\end{eqnarray} 
which satisfies the SUSY condition \eqn{cond non-compact spin(7)}. 
In this way we obtain superstring vacua of 
$Spin(7)$-holonomy manifolds. Simplest case $k=0$ again reduces 
to the model given in \cite{ES3}. Since $\dim G/H \neq 7$,
it seems difficult to relate these string vacua
with the solutions of SUGRA.

~

Before closing this section we want to make a few comments on the relation
between our CFT and the classical geometry of 
special holonomy manifolds based on the cone construction.
\begin{enumerate}
 \item The comparison between the geometrical cones 
and our ``CFT cones'' 
leads to an obvious disagreement  when $G/H$ is 
isomorphic with a round sphere $S^{9-d}$. 
For the $d=2$ string vacua, for example, all the cosets 
$G/H = SO(8)/SO(7)$, $SO(7)/G_2$, 
$SU(4)/SU(3)$, $SO(5)/SO(3)$ (the case when $SO(3)$ is embedded 
along a long root) are found to be topologically and metrically 
isomorphic with $S^7$. The cone over $S^7$ is of course 
the flat space $\br^8$, and hence allows the maximal spacetime SUSY.
On the other hand, the supercoset CFT's based on them are really inequivalent 
with each other, and yield string vacua with less spacetime SUSY.
As we discussed above, $SO(8)/SO(7)$ leads to 
non-SUSY vacua, and $SO(7)/G_2$, $SU(4)/SU(3)$, $SO(5)/SO(3)$
provide $Spin(7)$, $SU(4)$, $Sp(2)$ holonomies, respectively.

~

\item 
It is known that every 7-dim. tri-Sasakian manifold allows the second 
``squashed" Einstein metric that provides a weak $G_2$ holonomy \cite{FKMS,GS}
(see also \cite{BG,AFHS}).
A way to present the squashing procedure is to replace 
the original tri-Sasakian coset $G/H$ by 
$\dsp \frac{G\times SU(2)}{H\times SU(2)}$,
which is topologically isomorphic with $G/H$.   

For the case of $N(1,1) = SU(3)/U(1)$, the ``squashed supercoset CFT''
is defined in \eqn{squashed N11}. The similar construction is also 
possible for the tri-Sasakian coset $SO(5)/SO(3)$ \eqn{3S SO},
and could be identified as  the ``squashed $S^7$''
(often denoted as $J^7$).
Namely, we deform \eqn{3S SO} as 
\begin{eqnarray}
\cM&=& \frac{SO(5)_{k_1}\times SU(2)_{k_2}\times SO(7)_1}
{SO(3)_{\frac{k_1+1}{2}}\times SU(2)_{k_1+k_2+3}} ~,
\end{eqnarray}
where the $SU(2)$ in the denominator is embedded diagonally into 
$SU(2) \times SU(2) $, in which the first $SU(2)$-factor is the remaining 
one of $SO(5)/SO(3)$ and the second is the explicit $SU(2)$-factor.
We can easily show that it leads to string vacua with $Spin(7)$-holonomy
in the similar manner as in the $N(1,1)$ case, and the original tri-Sasakian
coset \eqn{3S SO} is recovered under the limit $k_2\,\rightarrow\,\infty$.

The analogous relation is also found in the $d=3$ example $S^3\times S^3$.
The coset CFT \eqn{S3 S3 3} may be regarded as the squashed version
of \eqn{S3 S3 1}. We again recover the unsquashed one \eqn{S3 S3 1} 
in the limit $k_3 \, \rightarrow\, \infty$.

\end{enumerate}

~

\section{Marginal Deformations: Spectrum of Cosmological Constant Operators}

\subsection{Cosmological Constant Operators Preserving Special Holonomy}

Since the linear dilation CFT is singular as a worldsheet theory, 
we should introduce 
the ``cosmological constant operators'' (Liouville potential terms) 
in order to eliminate its singular behavior at the tip of the cone 
(Liouville exponential prevents the field $\phi$ going out to $-\infty$). 
In this section we consider cosmological constant terms 
which preserve the spacetime supersymmetry and thus act as marginal perturbations 
in various models discussed in previous sections, focusing in particular
on the $G_2$ and $Spin(7)$ holonomy cases.

As in the old days of two-dimensional gravity \cite{2dgrav}, 
the cosmological constant operator is defined 
as the most relevant
primary field of the ``matter sector'' multiplied by the Liouville exponential. 
Here one might worry about the ``$\hat{c}=1$ ($c=3/2$) barrier'',
since the conformal matter $\br^{d-1,1} \times \cM$
has the central charge bigger than $3/2$ for any unitary $\cM$. 
In fact if one considers an identity operator (of matter sector) 
multiplied by a Liouville exponential, 
one finds the trouble of a complex Liouville exponent.  
This difficulty is avoided in our case by the 
requirement of GSO projection for vacua with unbroken spacetime SUSY.
We should define the cosmological constant term for the most relevant primary 
operator {\em allowed by the GSO condition}, and we can show that
the Liouville exponential is then always real as we shall 
see below. On the other hand 
in the case of broken spacetime SUSY, it seems difficult to define 
a suitable cosmological operator that resolves the
singularity, since the most relevant operator becomes tachyonic and 
has a complex Liouville exponential.

The general form of the marginal perturbation operator is written as follows;
\begin{align}
 & \left[G_{-\frac12}, \left[\bar G_{-\frac12}, 
\, e^{\gamma \phi}\Ocal^{(\sNS)}_{\cM} 
 \right]\right], 
\label{c t 1}
\end{align}
where $\Ocal^{(\sNS)}_{\cM}$ denotes an NS primary field in the $\cM$
sector. Here one can not choose the identity operator 
$\Ocal^{(\sNS)}_{\cM} = \id.$ since the operator (\ref{c t 1}) 
then becomes mutually non-local with respect to the spacetime SUSY operator (violates GSO
condition). The BRST invariance requires the on-shell condition 
\begin{eqnarray}
h(\Ocal^{(\sNS)}_{\cM}) + h(e^{\gamma \phi})
\equiv h(\Ocal^{(\sNS)}_{\cM}) -\frac{1}{2}\gamma^2
-\frac{1}{2}Q_{\phi}\gamma={1\over 2}~,
\end{eqnarray}
under which (\ref{c t 1}) manifestly preserves the worldsheet ${\cal N}=1$
supersymmetry.
To analyse the spectrum of operators $\Ocal^{(\sNS)}_{\cM}$ it is convenient to
consider their supersymmetric partners $\Ocal^{(\sR)}_{\cM}$ in the Ramond sector, which 
have the conformal weight
\begin{equation}
h(\Ocal^{(\sR)}_{\cM})=h(\Ocal^{(\sNS)}_{\cM})+{1-d \over 16}~.
\label{h OR ONS}\end{equation} 
This relation follows  from the spacetime supersymmetry:
Ramond states in the partition functions are dressed by
the spin fields of the Minkowski space and the Liouville fermion 
and possess the same conformal weights as the NS states. Dimensions 
of the spin fields add up to $(d-1)\over 16$ which accounts for the factor 
in (\ref{h OR ONS}).

The above relation may also be derived 
from the structure of our coset theories: 
NS and R states in the coset theory have the general form,
\begin{eqnarray}
&&\Ocal^{(\sNS)}_{\cM}=\Phi_{\Lambda,s=2,\lambda}
\left[{(G\times SO(9-d))/ H}\right]~,\nonumber \\
&&\Ocal^{(\sR)}_{\cM}=\Phi_{\Lambda,s=1(-1),\lambda}
\left[{(G\times SO(9-d))/ H}\right]~,
\label{nsr rep}\end{eqnarray}
where $\Phi_{\Lambda,s,\lambda}
\left[(G\times SO(9-d))/H\right]$ 
denotes a primary 
state in the coset $(G\times SO(9-d))/H$ defined by the 
highest weights $\Lambda$, $s$ and $\lambda$ of 
the affine Lie algebras $G$, $SO(9-d)$ and 
$H$, respectively.
$s=0,2,1,-1$ stand for the 
basic,vector, spinor and cospinor representations of $SO(9-d)$. 
Note that the weights $\Lambda,\lambda$ of NS and R states are the same for 
supersymmetric partners
and thus the difference in their dimensions come 
from that of the representations of
$SO(9-d)$. Difference of spinor and 
vector dimensions $(9-d)/16-1/2=(1-d)/16$ accounts
again for the RHS of (\ref{h OR ONS}). 
(In case a basic representation $s=0$ of the current algebra 
$SO(9-d)$ is used in the NS state, there is
an additional factor $+1/2$ in the RHS of (\ref{h OR ONS}).) 

The unitarity of $\cM$ sector requires the inequality 
$h(\Ocal^{(\sR)}_{\cM}) \geq c_{\cM}/24$, and thus we can easily show
that the Liouville exponent $\gamma$ 
is always real with the help of the condition \eqn{crit 2}. 

Let us now fix the value of the 
exponent $\gamma$ in the marginal perturbation operators (\ref{c t 1}).
From the old days of two-dimensional gravity it is known that 
Liouville exponentials have 
different characteristics depending on whether $\gamma>-Q_{\phi}/2$ 
or $\gamma < -Q_{\phi}/2$ \cite{Seiberg-L,GKP,Pelc} 
(see also \cite{GVW,ShV}). 
\begin{itemize}
 \item $ \gamma>-Q_{\phi}/2$~:~The operator $e^{\gamma \phi}$
       is called non-normalizable, 
       since the corresponding wave function exponentially diverges
       at the asymptotic region $\phi\rightarrow +\infty$. 
       It is interpreted as a coupling constant of the dual field theory,
       since the fluctuation has a divergent kinetic energy.
       In the context of two dimensional gravity it is also identified as the 
       local scaling operators. 
 \item $\gamma < -Q_{\phi}/2$~:~
        The operator $e^{\gamma \phi}$ is called normalizable.
        The wave function 
       is peaked around the singular region 
       $\phi \rightarrow -\infty$ as opposed to the above case.
       It is interpreted as a VEV of the dynamical fields
       of the dual theory (the modulus of vacuum),
       since the fluctuation has a finite kinetic energy.
\end{itemize}
Now we propose to choose the critical value for $\gamma$
\begin{equation} 
\gamma = -Q_{\phi}/2~,
\end{equation}
for our marginal operators (\ref{c t 1}).
 It corresponds to the maximal value
of conformal weight $h(e^{-\frac{Q_{\phi}}{2} \phi})= Q_{\phi}^2/8$ of the 
Liouville exponential with real $\gamma$ and corresponds
also to the minimum value in the continuous spectrum of delta-function
normalizable states $\gamma = -Q_{\phi}/2 +ip$ ($p \in \br$).
This situation is quite reminiscent of that of the $c=1$ conformal  
matter coupled to two dimensional gravity, or equivalently, 
the critical string on the background of 
two dimensional black hole \cite{2DBH}.

 It turns out that the condition $\gamma=-Q_{\phi}/2$ reduces the problem of 
finding marginal perturbations (\ref{c t 1}) to that of Ramond ground states in the theory $\cM$.
In fact when we take account of the ${\cal N}=1$ Liouville degrees of freedom and set
$\gamma=-Q_{\phi}/2$,
the Ramond sector
operator is given by
\begin{equation}
\cO^{(\sR)}=\sigma^{\phi}e^{-{Q_{\phi}\over 2}\phi}\cO^{(\sR)}_{\cM}~,
\label{ramondgr}\end{equation}
where $\sigma^{\phi}$ denotes the spin field 
associated with the Liouville fermion.
If we recall the criticality condition \eqn{crit 1}
\begin{equation}
{3\over 2}(d-2)+{3\over 2}+3Q_{\phi}^2+c_{\cM}=12~,
\end{equation}
and divide the formula by 24, we find
\begin{equation}
{1\over 16}+{Q_{\phi}^2\over 8}+{c_{\cM}\over 24}={10-d\over 16}~.
\end{equation}
A state with $h=(10-d)/16$ is in fact the Ramond ground state for the internal
space with $10-d$ dimensions. Thus the Ramond ground state of the system
$({\cal N}=1 \mbox{ Liouville})\times \cM$ 
is constructed from the Ramond ground state $h(\cO^{(\sR)}_{\cM})=c_{\cM}/24$
of the theory $\cM$.

We thus have an one-to-one correspondence between the marginal perturbation and
Ramond ground state as
\begin{equation}
\cO^{(\sNS)}=e^{-{Q_{\phi}\over 2}\phi}\cO^{(\sNS)}_{\cM}\Longleftrightarrow
\cO^{(\sR)}=\sigma_{\phi}e^{-{Q_{\phi}\over 2}\phi}\cO^{(\sR)}_{\cM}~.
\end{equation}
Here $\cO^{(\sNS)}_{\cM}$ and $\cO^{(\sR)}_{\cM}$ are related as (\ref{nsr rep}).
If one uses (\ref{h OR ONS}), one finds $h(\cO^{(\sNS)})=1/2$.
Such a correspondence between Ramond ground states and marginal operators was first
pointed out in \cite{SV}.
We will show below that in fact these operators $\cO^{(\sNS)}$ 
in the NS sector are marginal
perturbations preserving the special holonomies. We also note that
$\cO^{(\sNS)}_{\cM}$ 
is the most relevant primary field since the Liouville exponential 
$e^{-{Q_{\phi}\over 2}\phi}$ has the maximum dimension.

Our remaining task is to confirm that the cosmological constant operators 
defined here are really marginal deformations  preserving the spacetime SUSY.  
To this aim let us recall the discussions given in \cite{SV}. 
First of all,  the SCFT characterizing $G_2$ holonomy 
contains the tri-critical Ising model, 
and the energy momentum tensor
is decomposed as
\begin{align*}
 T=T^{\tri}+T^{r},\qquad T^{\tri}(z)T^{r}(w)\sim 0~,
\end{align*}
where $T^{\tri}$ and $T^{r}$ satisfy the Virasoro algebra with 
central charge $7/10$ and $49/5$. 
We express the conformal weights  as $(h^{\tri},h^{r})$
for $T^{\tri}$ and $T^{r}$ respectively. 
We only treat here operators
with the same right moving quantum numbers with the left moving ones, 
and focus only on the left movers.

As shown in \cite{SV}, the deformation \eqn{c t 1} 
preserves the spacetime supersymmetry and exactly marginal, 
if and only if $e^{-\frac{Q_{\phi}}{2} \phi}\Ocal^{(\sNS)}_{\cM}$ is a primary
of the type $(h^{\tri},h^{r})=\left(\frac{1}{10},\frac{2}{5}\right)$.
Its corresponding Ramond sector operator 
\begin{eqnarray}
\sigma^{\phi}_{\pm}e^{-\frac{Q_{\phi}}{2} \phi}\cO^{(\sR)}_{\cM}~,
\label{OR} 
\end{eqnarray}
(generated by the ``spectral flow operator'' 
$\left(\frac{7}{16},0\right)$) must then be an operator of the type 
$\left(\frac{3}{80},\frac{2}{5}\right)$.
The operator \eqn{OR} is doubly degenerate because of the spin field
$\sigma^{\phi}_{\pm}$, and we must take a suitable linear combination
compatible with the GSO projection, namely, 
the mutual locality with  spacetime supercharges.

In the \Spin(7) holonomy cases almost the same argument works. 
The energy momentum tensor can be decomposed into
the Ising part and the rest, and we write the conformal weights as 
 $(h^{\isi},h^{r})$. If and only if
$e^{-\frac{Q_{\phi}}{2} \phi}\Ocal^{(\sNS)}_{\cM}$ 
is a primary of the type 
$(h^{\isi},h^{r})=\left(\frac{1}{16},\frac{7}{16}\right)$, the deformation 
\eqn{c t 1} is an exactly marginal operator 
preserving $Spin(7)$-holonomy.
Moreover, such an NS primary corresponds to a Ramond state 
of the dimension $\left(\frac{1}{16},\frac{7}{16}\right)$.

Now, let us confirm  that these are indeed the cases for our cosmological 
constant operators \eqn{c t 1}.
In the $G_2$ holonomy case, the Ramond ground state operator must be either 
$\left(\frac{3}{80},\frac{2}{5}\right)$ or $\left(\frac{7}{16},0\right)$. 
In our construction $T^{\tri}$ is made up of sectors
$U(1)_{3/2} \times \psi^{\phi}$ and does not contain the Liouville field
$\phi$. We can hence conclude that $h^{r}\ne 0$ for the operator \eqn{OR}. 
Consequently the operator \eqn{OR} must be
of the $\left(\frac{3}{80},\frac{2}{5}\right)$ type.

In the \Spin(7) holonomy case, 
the problem is a little more subtle. There are three
types of Ramond ground states: $\left(0,\frac12\right),
\left(\frac12,0\right),\left(\frac{1}{16},\frac{7}{16}\right)$. 
Since $h^{r}\ne 0$ holds  by the same reason as $G_2$ holonomy case, 
the remaining possibilities are
$\left(0,\frac12\right)$ or $\left(\frac{1}{16},\frac{7}{16}\right)$.
Actually, the doubly degenerate
operators \eqn{OR} can be either of these types. 
However, it is possible to show that the operator 
$\left(\frac{1}{16},\frac{7}{16}\right)$, which has 
a mutually non-local OPE with the spectral flow operator
$\left(\frac12,0\right)$, survives after the GSO projection. 

In conclusion, the problem of enumerating marginal perturbations is
reduced to the classification of Ramond ground states of the conformal theory $\cM$
in the case of holonomies $G_2$ and $Spin(7)$.
It turns out that this also amounts to enumerating conformal 
blocks $F_2^{(*)}(h=1/2;\tau)$ defined in appendix B, appearing in the 
modular invariant partition functions. Each function in the NS sector
$F_2^{(\sNS)}(h=1/2;\tau)$ is one-to-one correspondence with
the operator $e^{-\frac{Q_{\phi}}{2}\phi}\cO_{\cM}^{(\sNS)}$
considered above.



We finally make a comment on the vacua with $\cN=2$ worldsheet SUSY;
\begin{eqnarray}
\br_{\phi} \times \cM \cong (\br_{\phi}\times S^1_Y) \times \cM/U(1)~,
\end{eqnarray}
where $\cM/U(1)$ is assumed to be an $\cN=2$ SCFT and 
$\br_{\phi}\times S^1_Y$ denotes the $\cN=2$ Liouville theory.
It is not difficult to show that the cosmological constant operator 
\eqn{c t 1} {\em does not\/} preserve the worldsheet $\cN=2$ SUSY.
The easiest way to see it is to observe that the integrality of 
the $U(1)_R$-charge fails for the operator of the type \eqn{c t 1}. 
Thus we now need to shift the exponent $\gamma$ away from $-Q_{\phi}/2$. 
Typical operators which preserve ${\cal N}=2$ SUSY are now
written as ;  
\begin{align}
 &\left[G_{-\frac12}^{-}, \left[\bar G_{-\frac12}^{-}, 
 \cO^{(\sNS)}_{\cM/U(1)} \exp\gamma\left(
 \phi+iY\right)\right]\right]+ \mbox{(c.c.)}~,  \nonumber \\
 &-\frac{Q_{\phi}}{2}\gamma+h(\cO^{(\sNS)}_{\cM/U(1)})=\frac12~,
 \label{c t 2}
\end{align}
where 
$\cO^{(\sNS)}_{\cM/U(1)}$ denotes a (cc) chiral primary field in 
the $\cM/U(1)$ sector. 
Again for the cases $\gamma > -Q_{\phi}/2$, \eqn{c t 2}
is non-normalizable and corresponds to a coupling constant,
while the operators $\gamma < -Q_{\phi}/2$ 
describe the  normalizable moduli that resolve the singularity.
The most relevant primary is of course $\cO^{(\sNS)}_{\cM/U(1)}=\id.$,
which corresponds to $\gamma = -1/Q_{\phi}$ and 
the well-known Liouville potential for the  $\cN=2$ Liouville theory.
In the case of singular Calabi-Yau $n$-folds, all of these 
operators \eqn{c t 2}
are normalizable for $n=2$ while non-normalizable for $n=4$.
In the $CY_3$ case, one ``half'' of them are normalizable and 
the remaining are non-normalizable. See for the detail 
\cite{GVW,GKP,ShV,Pelc,ES2}.

~

\subsection{Ramond Ground States in the $\cN=1$ Supercoset CFT's}

As we have shown, finding the possible marginal deformations
preserving $Spin(7)$ or $G_2$ holonomies amounts to the classification
of Ramond ground states in the supercoset theory $\cM$. 
We start by summarizing how to analyse this problem. 
It is quite reminiscent of the analysis on the chiral rings 
in the Kazama-Suzuki models \cite{LVW}.

Let us recall the structure of $\cN=1$ coset \eqn{N=1 coset}
and again assume $H=H_0 \times H_1 \times \cdots \times H_r$,
where $H_0$ denotes the abelian part and $H_i$ are simple parts.
For each $H_{i}$ ($i\neq 0$), 
the conformal dimension of the highest weight state 
$\lambda^{(i)}$ is evaluated as
\begin{eqnarray}
 h(\lambda^{(i)})&=&\frac{(\lambda^{(i)},\lambda^{(i)}+2\rho^{(i)})_{i}}
{2(k_i+h^{*}_i)}
=\frac{|\la^{(i)}+\rho^{(i)}|_i^2}{2(k_i+h^*_i)}+\frac{c_{H_i}}{24}
  -\frac{\dim H_i}{24}~ \nonumber \\
& =& \frac{|\la^{(i)}+\rho^{(i)}|^2}{2(k+g^*)}+\frac{c_{H_i}}{24}
  -\frac{\dim H_i}{24}~,
\end{eqnarray}
where $\rho^{(i)}$ denotes the Weyl vector of $H_{(i)}$ and 
$k_i$ is defined in \eqn{level formula}. The norm $|~|_i$ is associated with 
the inner product $(~,~)_i$.
We here made use of the well-known Freudenthal-de Vries strange formula:
$(\rho^{(i)},\rho^{(i)})_i= h_i^* \dim H_i/12$, and $c_{H_i} \equiv 
\frac{k_i\dim H_i}{k_i+h_i^*}$ denotes the central charge of $(H_i)_{k_i}$.
This formula also applies to the abelian part if we set $\rho^{(0)}=0$.
We thus find that 
the conformal dimension of the highest weight state
 $(\Lambda,s=1,\{\lambda^{(i)}\})$
of the Ramond sector becomes 
\begin{align}
 h(\Lambda,s=1,\{\lambda^{(i)}\})
  &= \frac{(\Lambda,\Lambda+2\rho_G)}{2(k+\hc)}+\frac{D}{16}
       -\sum_{i=0}^r 
  \frac{(\lambda^{(i)},\lambda^{(i)}+2\rho^{(i)})_i}{2(k_i+h_i^*)}\\
  &=
   \frac{|\Lambda+\rho_G|^2}{2(k+\hc)}
   -\sum_{i=0}^r\frac{|\lambda^{(i)}+\rho^{(i)}|^2}{2(k+\hc)}
       +\frac{c_{\Mcal}}{24}~,
\end{align}
where $\rho_G$ is the Weyl vector of $G$, $D\equiv \dim G/H$,
and $c_{\cM}$ is given in 
\eqn{c M}. 
This relation is valid only when the representation $\{\lambda^{(i)}\}$
is included in
the representation $\Lambda \times (\text{(co)spinor})$ as embedding
of finite dimensional Lie algebra $G\times \SO(D) \supset H$. 
Therefore we find the relation for the Ramond ground state
\begin{align}
 |\Lambda+\rho_G|^2 - |\lambda+\rho_H|^2=0~,
 \label{rgs-condition}
\end{align}
In this equation, we define $\lambda=\sum_i \lambda^{(i)}$
and $\rho_H=\sum_i \rho^{(i)}$. Formula (\ref{rgs-condition}) has been known in
${\cal N}=2$ coset theories \cite{LVW} and is now generalized to the case of ${\cal N}=1$ 
coset theories with spacetime supersymmetry.

Note that the condition (\ref{rgs-condition}) does not depend on the level $k$
of the affine algebra. 
The level $k$ enters only through the restriction on the possible 
set of representations $\Lambda$. Because of the unitarity,
the $\Lambda=\sum_{i}\Lambda_i \omega_i$ must satisfy the following relations.
\begin{align}
 \Lambda_i\in \Zb~,\quad \Lambda_i\ge 0~,\quad (\theta,\Lambda)\le k~.
 \label{integral-hw}
\end{align}
Only the finite number of $\Lambda$'s satisfy these relations.

In the next subsection we explicitly analyse the spectra of 
Ramond ground states. 
We treat all the cases of simple group $G$, and also the example 
$\cM= SU(2)^3/SU(2)$ of $G_2$ holonomy \eqn{S3 S3 3}.

~


\subsection{\Spin(7) Holonomy Cases}

We first study the Ramond ground states  
in the \Spin(7) holonomy cases. 
The wanted states are labeled by $(\Lambda,s=1,\lambda)$,
where $\Lambda$ is the integrable highest weight of $G_{k}$ and $\lambda$
is the integrable highest weight of $H$. 
We also use the label $\Lambda=\sum_{i}
\Lambda_i \omega_i$ and $\lambda=\sum_{i} \lambda_i \omega'_{i}$, where
$\omega_i$ and $\omega'_i$ are the fundamental weights of $G$ and $H$ 
respectively.

\begin{enumerate}
 \item \SO(7)/$G_2$: 
  The relevant supercoset is 
  \begin{align}
   \cM= &\frac{{\SO}(7)_{k} \times  {\SO}(7)_1}{( {G_2})_{k+1}}~.
  \end{align}
  Let us denote the highest weight of 
    $ {\SO}(7)_{k},  {\SO}(7)_1, ( {G_2})_{k+1}$   
    by $\Lambda=\sum_{j=1}^3\Lambda_j \omega_j,
    s=0,1,2,-1, \lambda=\sum_{j=1}^3\lambda_j \omega'_j$, respectively. 
    In this case, the Ramond ground states satisfy  
  \begin{align}
        \Lambda_3=2\Lambda_1+1~,\quad \lambda_1=\Lambda_2~,\quad 
    \lambda_2=\Lambda_1+\Lambda_3+1~,
  \end{align}
  in addition to the relations (\ref{integral-hw}). 
   The simplest example of the Ramond ground states is 
   $\Lambda_1=1,\Lambda_2=\Lambda_3=0$, which means $\Lambda$
    is the highest weight 
   of spinor representation of \SO(7) and $\lambda$ is the highest weight of
     $27$ dimensional representation of $G_2$.


 \item $SU(3)/U(1)$:
 The relevant supercoset is written as  
  \begin{align}
    \cM= \frac{\SU(3)_k\times \SO(7)_1}{\U(1)}~,
  \end{align}
 and the level of \U(1) is determined when we fix the embedding.
Consider the $N(m,\ell)$ type coset where \U(1) lies
   along the direction $\nu=(m-\ell)\omega_1+(m+2\ell)\omega_2$. We set
   $m$ and $\ell$ relatively prime, and $(m-\ell)\ge0, (m+2\ell) \ge 0$.
   If we use a canonically normalized \U(1) charge $p$ 
   (which means that the 
   dimension of an exponential operator becomes $p^2/2$), 
    the Ramond ground state 
   is obtained if it satisfies either of the following two conditions
  \begin{eqnarray}
   &&\hskip-10mm (a) \hskip3mm 
   \quad(m-\ell)(\Lambda_2+1)=(m+2\ell)(\Lambda_1+1)~,
   \quad 
   p=\frac{m(\Lambda_1+1)+(m+\ell)(\Lambda_2+1)}%
                 {\sqrt{2(k+3)(m^2+\ell^2-m\ell)}}~.\nonumber \\
  \\
   &&\hskip-10mm (b) \hskip3mm    
   \quad(m-\ell)(\Lambda_1+1)=(m+2\ell)(\Lambda_2+1)~,
   \quad 
   p=-\frac{m(\Lambda_2+1)+(m+\ell)(\Lambda_1+1)}%
                 {\sqrt{2(k+3)(m^2+\ell^2-m\ell)}}~.\nonumber \\
  \end{eqnarray}
As we already mentioned, we have the $\cN=4$ enhanced SUSY 
for the special case of $m=\ell=1$, and the $\cN=1$ cosmological terms 
\eqn{c t 1} cannot be applied in this case. We would like to 
further discuss this point elsewhere.

 \item \SO(5)/\SO(3)$_{\text{max}}$ :
  The relevant supercoset is expressed  as \eqn{SO(5)-SO(3) max};
  \begin{align}
     & \cM = \frac{ {\SO}(5)_{k} \times    
             {\SO}(7)_1}{ {\SU}(2)_{10k+28}}~.
   \end{align}
  If we denote by $(\ell+1)$ the the dimension of the representation of $SU(2)$,
  the Ramond ground state satisfies the relation
  \begin{align}
   \Lambda_2=2\Lambda_1+1~, \quad \ell=2\Lambda_1+5\Lambda_2+6~.
  \end{align}

  The number of the Ramond ground state is evaluated  as
  \begin{align*}
   \gauss{\frac{k+1}{3}}~,
  \end{align*}
where $\gauss{*}$ denotes the Gauss symbol.

  The simplest example of the Ramond ground state is 
  $\Lambda_1=0,\Lambda_2=1$, which means $\Lambda$ is the highest weight 
  of spinor representation of \SO(5) and $\lambda$ is the highest weight of
  $12$ dimensional representation of \SU(2).
\end{enumerate}

~


\subsection{$G_2$ Holonomy Cases}

We next analyse  the Ramond ground states for   
the $G_2$ holonomy case.
In these cases, the coset fermions 
$\SO(6)_1$ has two representations in Ramond sector:
spinor and cospinor. We express these representation by $s=\pm 1$ respectively.

\begin{enumerate}
\item $G_2$/\SU(3): 
The relevant supercoset is  given as 
\begin{align}
 \cM= \frac{(G_2)_k\times \SO(6)_1}{\SU(3)_{k+1}}~.
\end{align}
A Ramond ground state is obtained 
if either one of the following conditions is satisfied
\begin{eqnarray}
 && (a) \hskip3mm 
  s=-1~,\quad \lambda_1=\Lambda_1~,\quad \lambda_2=\Lambda_1+\Lambda_2+1~.\\
 && (b) \hskip3mm   s=1~,\quad \lambda_2=\Lambda_1~,\quad \lambda_1=\Lambda_1+\Lambda_2+1~.
 \end{eqnarray}
 The number of the Ramond ground states is evaluated  as follows;
\begin{align}
 \frac12(k+2)(k+3)-\left[\frac{k+2}{2}\right]~.
\end{align}
The simplest Ramond ground states come from the basic representation of $G_2$,
and represented as $(\Lambda=0,s=-1,\lambda_1=0,\lambda_2=1)$ and 
$(\Lambda=0,s=+1,\lambda_1=1,\lambda_2=0)$.


  \item \SO(5)/(\SU(2) $\times$ \U(1)):
The relevant coset corresponds to the (non-HSS)
$\CP^3$ \eqn{CP3 G2};
\begin{align}
 \cM= \frac{\SO(5)_k\times \SO(6)_1}{\SU(2)_{k+1}\times \U(1)_{k+3}}~.
\end{align}
 We denote by $(\ell+1)$ the dimension of the representation of
 \SU(2), and by $m$ the charge of \U(1) normalized 
so that
 the conformal dimension 
of its vertex operator is given by $\frac{m^2}{4(k+3)}$.
 In this notation, a Ramond ground state is obtained 
if it satisfies one of the
 following four conditions
\begin{eqnarray}
    && (a) \hskip3mm    
s=-1~,\quad \ell=\Lambda_1~,\quad m=\Lambda_1+\Lambda_2+2~.\\
    && (b) \hskip3mm   
s=+1~,\quad \ell=\Lambda_1+\Lambda_2+1~,\quad m=\Lambda_1+1~. \\
    && (c) \hskip3mm     
s=-1~,\quad \ell=\Lambda_1+\Lambda_2+1~,\quad m=-\Lambda_1-1~. \\
    && (d) \hskip3mm     
s=+1~,\quad \ell=\Lambda_1~,\quad m=-\Lambda_1-\Lambda_2-2~.
\end{eqnarray}
As a result, there are four Ramond ground states for each representation 
$\Lambda$ of \SO(5). However there are the field identifications
originating from the outer automorphism $\Zb_2$;
\begin{align}
 (\Lambda_1,\Lambda_2,s,\ell,m)
\cong(k-\Lambda_1-\Lambda_2,\Lambda_2,s+2,
  k-\ell,m+(k+3))~.
\end{align}
Hence, the number of Ramond ground states is given by
\begin{align}
 (k+1)(k+2)~.
\end{align}
Simplest ground states are comes from the basic representation of
\SO(5): $(\Lambda=0,s=1,\ell=1,m=1)$ and $(\Lambda=0,s=1,\ell=0,m=-2)$ .


\item $\SU(3)/\U(1)^2$ : The relevant coset is characterized by 
the charge lattice $\Gamma$ of $U(1)^2$. We here only consider 
the simplest case $\Gamma = \bz \nu_1+ \bz\nu_2$, where 
$\nu_1$ and $\nu_2$ are defined in \eqn{identification2}, 
leading to
\begin{align}
 \cM= \frac{\SU(3)_{k}\times \SO(6)_1}{U(1)_{k+3}\times U(1)_{3(k+3)}}~.
\end{align}
We label the $\U(1)$ charge by 
$\lambda=\lambda_1\omega_1+\lambda_2\omega_2$
normalized so that the dimension of the vertex operator is given by 
$\frac{(\lambda,\lambda)}{2(k+3)}$.
The Ramond ground states satisfy the condition
\begin{align}
 w(\Lambda+\rho)=\lambda~,\quad s=-\epsilon(w)~,
\end{align}
for an element $w$ of the $\SU(3)$ Weyl group, where $\epsilon(w)$ is the 
signature of $w$.
We here note a slightly non-trivial point. The charge lattice $\Gamma$
here is {\em not\/} invariant under the outer automorphism $\bz_3$ 
which usually implies the necessity of field identification. 
We thus should consider the spectrum without field identification
contrary to the standard Kazama-Suzuki coset $\Gamma = \sqrt{k+3}Q$.  
Consequently, the number of the Ramond ground states becomes 
\begin{align}
 3(k+1)(k+2)~.
\end{align}
The simplest Ramond ground states are expressed as
$(\Lambda=0,s=-1,\lambda_1=1,\lambda_2=1)$ and its Weyl transforms.


 \item \SU(2)${}^{3}$/\SU(2):
The supercoset is expressed as \eqn{S3 S3 3};
\begin{align}
 & \cM= \frac{ {\SU}(2)_{k_1}\times  
{\SU}(2)_{k_2} \times  {\SU}(2)_{k_3}  \times  {\SO}(6)_1}{
 {\SU}(2)_{k_1+k_2+k_3+4}}~.
\end{align}
Let us denote 
the dimension of representation by $\ell_j+1,\ (j=1,2,3)$
for each \SU(2) factor in the numerator,
and the one in the denominator by $\ell_4+1$
One obtains a Ramond ground state if
\begin{align}
 \frac{\ell_1+1}{k_1+2}=\frac{\ell_2+1}{k_2+2}=\frac{\ell_3+1}{k_3+2}~,\nn\\
 \ell_4=\ell_1+\ell_2+\ell_3+3~,\quad s=\pm 1~.
\end{align}
are  satisfied. In this case, by
denoting the greatest common divisor of $\{k_j+2\}$ as $p$,
the number of Ramond ground states becomes equal to 
$(p-1)$,  when we take a suitable field identification into account.

\end{enumerate}

\section{Discussions}

In this paper we have investigated aspects of superstring vacua 
of the type: 
$$
\br^{d-1,1}\times (\cN=1 ~ \mbox{Liouville}) \times 
(\cN=1 ~ \mbox{supercoset CFT}~ \mbox{on } G/H)~,
$$
motivated by an analogy with the geometrical cone constructions of
special holonomy manifolds over the Einstein homogeneous spaces $G/H$.
We made an almost exhaustive analysis and obtained results which 
led in most cases to the same amount of supersymmetries
as expected from the geometrical approach.

While these results seem satisfactory, 
we should also emphasize the obvious difference between our and geometrical approaches. 
In the non-linear $\sigma$-model on the geometrical cone over $G/H$,
the physics depends on the choice of its metric, and possibly
on the global topology. On the other hand, 
in our algebraic approach based on the coset CFT's, the vacua 
are defined associated with the affine Lie algebras for $G$, $H$
and further with the choice of embedding of $\mbox{Lie}\,(H)$ 
into $\mbox{Lie}\,(G)$. We do not use information on the metric structure
and global topology of the manifold $G/H$ explicitly. (It will be interesting to check 
if one obtains metrics used in the geometrical cone constructions when one
computes the natural metric on homogeneous space making use of our embedding 
of $H$ into $G$ described in various examples).
Actually we often encountered examples 
where the same manifold has several different coset realizations.
They are of course equivalent in the sense of non-linear $\sigma$-model, 
but {\em not\/} necessarily equivalent as the coset CFT's.  
The typical example is the case of round sphere. 
The geometrical cone over a round sphere is a trivial flat space and 
leads to a vacua with a maximal amount of SUSY, while we have several inequivalent
coset CFT's based on a round sphere
possessing different amount of supercharges.

Important questions for our analysis may as follows:
\begin{enumerate}
 \item How can we identify our string vacua
       with the known solutions in supergravity theories?
 \item How can we attach a physical meaning to the levels of 
       current algebras in our construction?
 \item What is the precise geometrical interpretation of 
       the Ramond ground states discussed in section 4? 
\end{enumerate} 
These problems are deeply related with each other and seem 
difficult in particular in the cases of $G_2$
and $Spin(7)$ holonomies.

However, we now would like to point out some suggestive examples: 
For the $d=4$ vacua with $\cM= (SU(2)_{k_1}\times SU(2)_{k_2})/U(1)$, 
we observed that the spacetime SUSY is allowed only for the case of
$U(1)$ embedding $p=q$ in \eqn{gamma Tpq} which corresponds to the Einstein homogeneous 
space $T^{1,1}$.
Moreover, it is easy to see that the simplest case $k_1=k_2=0$ is
equivalent to the string theory on conifold as discussed in \cite{GV}. 
The brane 
construction of conifold was explored in \cite{DM} based on 
the ALE fibration and the well-known ALE-NS5 correspondence \cite{OV,GHM}.
It is realized as a system of two intersecting NS5-branes, 
both of which fill the 4-dimensional Minkowski spacetime.
Because the level of WZW-model is translated into
the brane charge of NS5, it is natural 
to suppose that the higher level models 
$\cM = (SU(2)_{k_1}\times SU(2)_{k_2})/U(1)$ 
could be associated with a configuration of intersecting 
stacks of NS5-branes with brane charges $k_1$ and $k_2$.  
In fact, it is known \cite{GKP,HK} that the vacua 
with $k_1=k$, $k_2=0$ are understood as the ADE hierarchy 
of rational singularities, or the wrapped NS5
branes with charge $\simeq k$. 

On the other hand, the brane interpretation for the known supergravity 
solutions of $G_2$ (and $Spin(7)$) holonomy have been discussed 
in \cite{AW,GT}. In \cite{AW} the relevant brane configurations
are identified as $D6$-brane wrapped around special Lagrangian cycles  
(``L-pictures''), which are again reinterpreted as the 
intersecting NS5-branes by some duality web   
\cite{GT}. The cases of intersecting stacks of NS5-branes
still describe the vacua with the correct number of spacetime SUSY's, 
although  the explicit supergravity solutions are not available.
It seems again plausible to assume that the charges of
stacked  NS5-branes are incorporated as the levels of WZW models in our construction.

Encouraged by these observations, 
we conjecture for the string vacua with 
$G_2$ and $Spin(7)$ holonomies;
\begin{enumerate}
 \item Our coset CFT models for the nearly K\"{a}hler (weak $G_2$)
       homogeneous spaces $G_k/H_{*}$ 
      should be identified as the geometrical $G_2$ 
       ($Spin(7)$) cones over them 
      {\em in the special case of zero level $k=0$.}
 \item The non-zero level models could be associated with the same 
       configurations of intersecting NS5-branes as for the geometrical 
       $G_2$ ($Spin(7)$) cones considered above, with the 
       suitable numbers of stacked NS5-branes.
 \item Marginal deformations discussed in section 4 correspond
       to various motions of 
       intersecting NS5-branes preserving the spacetime SUSY.
\end{enumerate}

In this paper we have assumed the special value of the Liouville momentum
$\gamma=-Q_{\phi}/2$ and have constructed associated marginal 
perturbation operators.
An obvious question is if one can construct other marginal operators
with a different choice for the momentum. It seems that this is a problem whose
answer depends to some extent on how one defines the Liouville theory. 
As is well-known there are two branches for the Liouville momentum
\begin{align}
&&&\mbox{(i)  \hskip3mm     ``discrete series"}:  \hskip7mm &&\gamma \in \br \\
&&&\mbox{(ii) \hskip3mm     ``principal continuous series"}: \hskip7mm 
&&\gamma=-{Q_{\phi}\over 2}+ip~, \quad p \in \br
\end{align}
and our choice $-Q_{\phi}/2$ was special in the sense that with this value of
$\gamma$ the dimension of the Liouville exponential $e^{\gamma\phi}$
is maximum for the series (i) and minimum for the series (ii). Relevant
range of the ``discrete series" here is $\gamma<-Q_{\phi}/2$ which provides 
operators peaked around the tip of the cone and possibly resolve its singular 
behavior.

When one takes the Liouville momentum from the ``discrete series" 
$\gamma<-Q_{\phi}/2$,  the dimension $h(e^{\gamma\phi})$
could become negative and large as $-\gamma$ becomes large. 
One may possibly construct marginal operators by pairing such an exponential 
with a field from the matter 
sector $\cM$ with a large positive dimension. 
However, such a possibility seems 
rather unphysical since the Liouville exponential with a large negative 
$\gamma$ should effect strongly the physics around $\phi \approx -\infty$
while matter fields of high dimensions should be largely 
irrelevant.

Unitarity of the Liouville theory is a difficult and subtle
problem, however, it seems generally agreed that the theory becomes unitary 
when one restricts oneself to the sector of ``principal continuous
 series". If one takes 
this point of view, the value $\gamma=-Q_{\phi}/2$ appears to be the unique
choice for marginal operators since this is the value for which the
operator $e^{\gamma\phi}$ is real and has the lowest dimension.
In fact the central charge of the ${\cal N}=1$ Liouville system is given by
$c_L=3/2+3Q_{\phi}^2$ and the operator
$\sigma^{\phi}e^{{-Q_{\phi}\over 2}\phi}$ saturates its BPS bound
$c_L/24=1/16+Q_{\phi}^2/8$.    

Thus we further conjecture 
\begin{enumerate}
\setcounter{enumi}{3}
 \item  Marginal operators we have constructed exhaust all possible marginal 
perturbations
of the ${\cal N}=1$ coset model of string vacua for manifolds with special
holonomy.
\end{enumerate}

In order to discuss these conjectures it is quite important 
to establish the dictionary translating the cosmological constant  
operators into the geometrical data. To this aim, a possible future direction
may be the boundary state analysis along the line similar to
\cite{ES2}. It may be also interesting and challenging to try to
generalize the concept of chiral rings characteristic for 
the $\cN=2$ string vacua \cite{LVW}. It is quite suggestive  
that our cosmological constant operators are defined 
in one-to-one correspondence  with the Ramond ground states
for  the ``base'' $\cM$, implying some cohomological structure behind 
the system. The approach based on the {\em real\/} Landau-Ginzburg 
theory may also be significant.

~


\subsection*{Acknowledgements}
\indent
We would like to thank  G. M. Sotkov for informing us 
the work \cite{Sotkov} after our previous paper \cite{ES3} was
published. We also would like to thank H. Kanno for useful information on
special holonomy cones.

Research of T.E. and Y.S. is supported in part by a Grant-in-Aid  
from Japan Ministry of Education, Culture, Sports, Science and Technology. Research of S.Y. 
is supported in part by Soryushi Shogakukai.


\section*{Appendix A : ~ Notations}
\setcounter{equation}{0}
\def\theequation{A.\arabic{equation}}
\indent

 Set $q:= e^{2\pi i \tau}$ , $y:=e^{2\pi i z}$ ;
 \begin{equation}
 \begin{array}{l}
 \dsp \th_1(\tau,z) =i\sum_{n=-\infty}^{\infty}(-1)^n q^{(n-1/2)^2/2} y^{n-1/2}
  \equiv 2 \sin(\pi z)q^{1/8}\prod_{m=1}^{\infty}
    (1-q^m)(1-yq^m)(1-y^{-1}q^m)~, \\
 \dsp \th_2(\tau,z)=\sum_{n=-\infty}^{\infty} q^{(n-1/2)^2/2} y^{n-1/2}
  \equiv 2 \cos(\pi z)q^{1/8}\prod_{m=1}^{\infty}
    (1-q^m)(1+yq^m)(1+y^{-1}q^m)~, \\
 \dsp \th_3(\tau,z)=\sum_{n=-\infty}^{\infty} q^{n^2/2} y^{n}
  \equiv \prod_{m=1}^{\infty}
    (1-q^m)(1+yq^{m-1/2})(1+y^{-1}q^{m-1/2})~, \\
 \dsp \th_4(\tau,z)=\sum_{n=-\infty}^{\infty}(-1)^n q^{n^2/2} y^{n}
  \equiv \prod_{m=1}^{\infty}
    (1-q^m)(1-yq^{m-1/2})(1-y^{-1}q^{m-1/2})~.
 \end{array}
 \end{equation}
 \begin{eqnarray}
 \Th{m}{k}(\tau,z)&=&\sum_{n=-\infty}^{\infty}
 q^{k(n+\frac{m}{2k})^2}y^{k(n+\frac{m}{2k})} ~,\\
 \tTh{m}{k}(\tau,z)&=&\sum_{n=-\infty}^{\infty} (-1)^n
 q^{k(n+\frac{m}{2k})^2}y^{k(n+\frac{m}{2k})}~.
 \end{eqnarray}
 We often use the abbreviations; $\th_i \equiv \th_i(\tau, 0)$
 ($\th_1\equiv 0$), $\Th{m}{k}(\tau) \equiv \Th{m}{k}(\tau,0)$ ,
 $\tTh{m}{k}(\tau) \equiv \tTh{m}{k}(\tau,0)$ .
 We also set
 \begin{equation}
 \eta(\tau)=q^{1/24}\prod_{n=1}^{\infty}(1-q^n) ~.
 \end{equation}
The product formula of theta functions is useful for our analysis;
 \begin{equation}
 \Th{m}{k}(\tau,z)\Th{m'}{k'}(\tau,z')
 =\sum_{r\in\bz_{k+k'}}\Th{mk'-m'k+2kk'r}{kk'(k+k')}(\tau,u)
 \Th{m+m'+2kr}{k+k'}(\tau,v)~,
 \label{product}
 \end{equation} 
 where we set 
 $\dsp u= \frac{z-z'}{k+k'}$, $\dsp v=\frac{kz+k'z'}{k+k'}$.

The next formula is also useful
\begin{equation}
\Th{m/p}{k/p}(\tau,z)=\Th{m}{k}(\tau/p,z/p)=
\sum_{r\in\bz_p} \,\Th{m+2kr}{pk}(\tau,z/p)~,
\label{theta identity}
\end{equation}
where $m$, $k$ are some real numbers and $p$ is an integer.

~

\section*{Appendix B : ~ Massive Characters of the Extended Chiral
Algebras Associated to Special Holonomies}
\setcounter{equation}{0}
\def\theequation{B.\arabic{equation}}
\indent

Here we summarize the massive character formulas 
of the extended chiral algebras characterizing the special 
holonomy manifolds. For simplicity we shall only focus on the NS sector.
The characters for other spin structures are obtained by making the
half-integral spectral flows as follows;
\begin{eqnarray}
\Ch{(\stNS)}(*;\tau,z)&=& \Ch{(\sNS)}(*;\tau,z+\frac{1}{2})~, \nonumber \\
\Ch{(\sR)}(*;\tau,z)&=& q^{\frac{c}{24}}y^{\frac{c}{6}}
\Ch{(\sNS)}(*;\tau,z+\frac{\tau}{2}) ~, \nonumber \\
\Ch{(\stR)}(*;\tau,z)&=& q^{\frac{c}{24}}y^{\frac{c}{6}}
\Ch{(\sNS)}(*;\tau,z+\frac{\tau}{2}+\frac{1}{2}) ~.
\end{eqnarray}

\noindent
{\bf $1. ~ Sp(k)$\mbox{-holonomy}  $(k=1,2)$ } :

This case corresponds to hyper K\"{a}hler manifolds of the real 
dimension $4k$.
The relevant chiral algebra is the (small) $\cN=4$
superconformal algebra with the level $k$ ($c=6k$), as is well-known.
The massive representations are labeled by the conformal weight $h$
and the $SU(2)$ spin $\ell/2$, and the unitarity requires the
constraints $h\geq \ell/2$.
The character formulas are given in \cite{ET};
\begin{eqnarray}
\Ch{(\sNS)}(h,\ell;\tau,z) &=& 
\frac{q^{h-\frac{(\ell+1)^2}{4(k+1)}-\frac{1}{4}}}
{\eta(\tau)} \left(\frac{\th_3(\tau,z)}{\eta(\tau)}\right)^2
\chi_{\ell}^{(k-1)}(\tau,z) ~,~~~(0\leq \ell \leq k-1)~,
\end{eqnarray}
where $\chi^{(k)}_{\ell}(\tau, z)$ denotes the 
character of $SU(2)_k$ with the spin $\ell/2$ ($0\leq \ell \leq k$)
representation;
\begin{equation}
\chi^{(k)}_{\ell}(\tau, z) =\frac{\Th{\ell+1}{k+2}-\Th{-\ell-1}{k+2}}
                        {\Th{1}{2}-\Th{-1}{2}}(\tau, z) ~  .
\end{equation}
We note that
\begin{eqnarray}
\chi^{(1)}_0(\tau,z) &=& \frac{\Th{0}{1}(\tau,2z)}{\eta(\tau)} \nonumber \\
\chi^{(1)}_1(\tau,z) &=& \frac{\Th{1}{1}(\tau,2z)}{\eta(\tau)}~.
\end{eqnarray}

~

\noindent
{\bf $2.~SU(n)$-holonomy}  : 

This case corresponds to the Calabi-Yau $n$-fold ($CY_n$) compactification.
The extended chiral algebras are defined by adding the integral 
spectral flow operator, which corresponds to the 
holomorphic $n$-form in $CY_n$, to the $\cN=2$
superconformal algebra with $c=3n$. These conformal algebras 
are not Lie algebras but are W-algebras except for the
$SU(2)$-holonomy case, which of course reduces to the above
$Sp(1)$ case. For the $SU(3)$-holonomy the spectral flow operator
generates the subsector described by $SO(6)_1/SU(3)_1\cong U(1)_{3/2}$,
which is equivalent with the $\cN=2$ minimal model of level 1.  
The corresponding subsector for the $SU(4)$-holonomy 
is $SO(8)_1/SU(4)_1 \cong U(1)_2$, which is equivalent with 
the conformal system of a complex fermion.

The massive representations are labeled 
by the conformal weight $h$ and the $U(1)_R$ charge $Q$
with the unitarity condition $h\geq Q/2$.
The character formulas 
for $SU(3)$-holonomy are given in \cite{odake}, and those for 
$SU(4)$-holonomy are given in \cite{HS}.
\begin{itemize}
 \item $SU(3)$-holonomy : 
We have two continuous series;
\begin{eqnarray}
\Ch{(\sNS)}(h,Q=0;\tau,z) &=& \frac{q^{h-\frac{1}{4}}}{\eta(\tau)}
\frac{\th_3(\tau,z)}{\eta(\tau)}\frac{\Th{0}{1}(\tau,2z)}{\eta(\tau)} ~,
\nonumber \\
\Ch{(\sNS)}(h,|Q|=1;\tau,z) &=& \frac{q^{h-\frac{1}{2}}}{\eta(\tau)}
\frac{\th_3(\tau,z)}{\eta(\tau)}\frac{\Th{1}{1}(\tau,2z)}{\eta(\tau)} ~.
\end{eqnarray}
The vacuum state is doubly degenerate for the second case ($Q=1$ and $Q=-1$).
 \item $SU(4)$-holonomy :
We have three continuous series;
\begin{eqnarray}
\Ch{(\sNS)}(h,Q=0;\tau,z) &=& \frac{q^{h-\frac{3}{8}}}{\eta(\tau)}
\frac{\th_3(\tau,z)}{\eta(\tau)}\frac{\Th{0}{\frac{3}{2}}
(\tau,2z)}{\eta(\tau)} ~,
\nonumber \\
\Ch{(\sNS)}(h,Q=\pm 1;\tau,z) &=& \frac{q^{h-\frac{13}{24}}}{\eta(\tau)}
\frac{\th_3(\tau,z)}{\eta(\tau)}\frac{\Th{\pm 1}{\frac{3}{2}}
(\tau,2z)}{\eta(\tau)} ~.
\end{eqnarray}
\end{itemize}

~

\noindent
{\bf $3. ~ G_2~\mbox{and}~ Spin(7)$-holonomies} : 

The chiral algebras associated to the $G_2$ and $Spin(7)$-holonomies
are again the W-algebra like extensions of $\cN=1$ superconformal
algebra explicitly defined in \cite{SV}\footnote
   {In some literature the $G_2$ extended algebra is denoted as 
    ${\cal SW}(3/2,3/2,2)$-algebra ($c=21/2$) and the $Spin(7)$
     extended algebra is done as ${\cal SW}(3/2,2)$-algebra ($c=12$).}. 
A characteristic feature of the $G_2$ extended algebra 
is the existence of the tri-critical Ising  
model ($\cong SO(7)_1/(G_2)_1$), and that for the $Spin(7)$ extended 
algebra is the Ising model ($\cong SO(8)_1/SO(7)_1$) as discussed 
in \cite{SV}. The unitary massive representations for these algebras 
are classified in \cite{GN,Noyvert}: two continuous series with $h\geq 0$ and 
$h\geq 1/2$ exist for the each case. Unfortunately,
the character formulas for these representations
have not been worked out. 
Nevertheless, based on the existence of
spacetime SUSY as well as the worldsheet
$\cN=1$ superconformal symmetry, it seems plausible to
propose that the conformal blocks corresponding to the 
massive representations should be expanded  
into the following functions:
\begin{itemize}
 \item $Spin(7)$-holonomy~: For the NS sector,
\begin{eqnarray}
F^{(\sNS)}_1(h;\tau)&=& \frac{q^{h-\frac{49}{120}}}{\eta(\tau)}
\sqrt{\frac{\th_3}{\eta}}(\tau)\chi_0^{\msc{tri},\, (\sNS)}(\tau) \nonumber\\
&\equiv& \frac{q^{h-\frac{49}{120}}}{\eta(\tau)} \left(
\chi^{\msc{Ising}}_0(\tau)\chi^{\msc{tri}}_0(\tau)
+\chi^{\msc{Ising}}_{1/2}(\tau)\chi_{3/2}^{\msc{tri}}(\tau)
+\chi^{\msc{Ising}}_{1/16}(\tau)\chi^{\msc{tri}}_{7/16}(\tau) 
\right) ~,\nonumber \\
&& \hspace{9cm} (h\geq 0) \nonumber \\
F^{(\sNS)}_2(h;\tau)&=& \frac{q^{h-\frac{1}{10}-\frac{49}{120}}}{\eta(\tau)}
\sqrt{\frac{\th_3}{\eta}}(\tau)\chi_{1/10}^{\msc{tri},\, (\sNS)}(\tau) 
\nonumber\\
&\equiv& \frac{q^{h-\frac{1}{10}-\frac{49}{120}}}{\eta(\tau)} \left(
\chi^{\msc{Ising}}_0(\tau)\chi^{\msc{tri}}_{3/5}(\tau)
+\chi^{\msc{Ising}}_{1/2}(\tau)\chi_{1/10}^{\msc{tri}}(\tau)
+\chi^{\msc{Ising}}_{1/16}(\tau)\chi^{\msc{tri}}_{3/80}(\tau) 
\right)~, \nonumber \\
&& \hspace{9cm} (h\geq 1/2)
\label{F NS spin(7)}
\end{eqnarray}
where we have written $\chi^{\msc{tri},\,(\sNS)}_h$, $\chi^{\msc{tri}}_h$ for 
the $\cN=1$ and $\cN=0$ characters of the tri-critical Ising model, 
and $\chi^{\msc{Ising}}_h$ for the Ising model. 
The second lines are consistent with the decompositions of massive 
representations of the $Spin(7)$ extended algebra with respect to the 
Virasoro modules of Ising model presented in \cite{GN}.
This fact suggests that the functions 
$F^{(\sNS)}_i(h;\tau)$ may in fact be the massive characters 
although we have not been able to prove this.

\item $G_2$-holonomy~: For the NS sector, the wanted functions are given as
\begin{eqnarray}
F^{(\sNS)}_1(h;\tau)&=& \frac{q^{h-\frac{1}{3}}}{\eta(\tau)}
\sqrt{\frac{\th_3}{\eta}}(\tau)\frac{\Th{0}{3/2}}{\eta(\tau)} 
\nonumber \\
&\equiv & \frac{q^{h-\frac{1}{3}}}{\eta(\tau)} 
\left(\chi^{\msc{tri},\,(\sNS)}_0 \chi^{\msc{Potts}}_0 +
\chi^{\msc{tri},\,(\sNS)}_{1/10} \chi^{\msc{Potts}}_{2/5}
\right) (\tau) ~, ~~~(h\geq 0) \nonumber \\
F^{(\sNS)}_2(h;\tau)&=& \frac{q^{h-\frac{1}{2}}}{\eta(\tau)}
\sqrt{\frac{\th_3}{\eta}}(\tau)\frac{\Th{1}{3/2}}{\eta}(\tau) \nonumber \\
&\equiv & \frac{q^{h-\frac{1}{2}}}{\eta(\tau)} 
\left(\chi^{\msc{tri},\,(\sNS)}_0 \chi^{\msc{Potts}}_{2/3} +
\chi^{\msc{tri},\,(\sNS)}_{1/10} \chi^{\msc{Potts}}_{1/15}
\right) (\tau) ~, ~~~ (h\geq 1/2)
\label{F NS g2}
\end{eqnarray}
where $\chi^{\msc{tri},\,(\sNS)}_h$ again means the $\cN=1$ character
of tri-critical Ising and $\chi^{\msc{Potts}}_h$ denotes the character 
of the 3-state Potts model ($\bz_3$ Parafermion) that is defined as the 
coset CFT $SU(2)_3/U(1)_3$. The second lines are easily derived  using the 
equivalence 
\begin{eqnarray}
SO(1)_1 \times \frac{SO(6)_1}{SU(3)_1}& \cong& \frac{SO(7)_1}{(G_2)_1} \times 
\frac{(G_2)_1}{SU(3)_1}~ \nonumber \\
&\cong & \mbox{tri-critical Ising} \times \mbox{3-state Potts}~.
\end{eqnarray}
They are consistent with 
the structures of unitary massive representations given in \cite{Noyvert}.

\end{itemize}

\newpage



\begin{thebibliography}{100}


\bibitem{Strominger}
A.~Strominger,
Nucl.\ Phys.\ B {\bf 451}, 96 (1995)
[arXiv:hep-th/9504090];
B.~R.~Greene, D.~R.~Morrison and A.~Strominger,
Nucl.\ Phys.\ B {\bf 451}, 109 (1995)
[arXiv:hep-th/9504145].


\bibitem{Aspinwall}
P.~S.~Aspinwall,
Phys.\ Lett.\ B {\bf 357}, 329 (1995)
[arXiv:hep-th/9507012].


\bibitem{Witten}
E.~Witten,
{\it ``Some comments on string dynamics,''}
arXiv:hep-th/9507121.



\bibitem{AMV}
M.~Atiyah, J.~M.~Maldacena and C.~Vafa,
J.\ Math.\ Phys.\  {\bf 42}, 3209 (2001)
[arXiv:hep-th/0011256].


\bibitem{AW}
M.~Atiyah and E.~Witten,
arXiv:hep-th/0107177.


\bibitem{AdSCFT}
J.~M.~Maldacena,
Adv.\ Theor.\ Math.\ Phys.\  {\bf 2}, 231 (1998)
[Int.\ J.\ Theor.\ Phys.\  {\bf 38}, 1113 (1999)]
[arXiv:hep-th/9711200];
S.~S.~Gubser, I.~R.~Klebanov and A.~M.~Polyakov,
Phys.\ Lett.\ B {\bf 428}, 105 (1998)
[arXiv:hep-th/9802109];
E.~Witten,
Adv.\ Theor.\ Math.\ Phys.\  {\bf 2}, 253 (1998)
[arXiv:hep-th/9802150];
O.~Aharony, S.~S.~Gubser, J.~Maldacena, H.~Ooguri and Y.~Oz,
Phys.\ Rept.\  {\bf 323} (2000) 183,
hep-th/9905111. 



\bibitem{KW}
I.~R.~Klebanov and E.~Witten,
Nucl.\ Phys.\ B {\bf 536}, 199 (1998)
[arXiv:hep-th/9807080].


\bibitem{AFHS}
J.~M.~Figueroa-O'Farrill,
{\it ``Near-horizon geometries of supersymmetric branes,''}
arXiv:hep-th/9807149;
B.~S.~Acharya, J.~M.~Figueroa-O'Farrill, C.~M.~Hull and B.~Spence,
Adv.\ Theor.\ Math.\ Phys.\  {\bf 2}, 1249 (1999)
[arXiv:hep-th/9808014];
J.~M.~Figueroa-O'Farrill,
Class.\ Quant.\ Grav.\  {\bf 16}, 2043 (1999)
[arXiv:hep-th/9902066].



\bibitem{Gubser}
S.~S.~Gubser,
Phys.\ Rev.\ D {\bf 59}, 025006 (1999)
[arXiv:hep-th/9807164].

\bibitem{MP}
D.~R.~Morrison and M.~R.~Plesser,
Adv.\ Theor.\ Math.\ Phys.\  {\bf 3}, 1 (1999)
[arXiv:hep-th/9810201].





\bibitem{Sasaki}
S. Sasaki, 
T$\hat{\mbox{o}}$hoku Math. J. {\bf 2} (1960), 459-476.



\bibitem{BG}
C.~P.~Boyer and K.~Galicki,
Surveys Diff.\ Geom.\  {\bf 7}, 123 (1999)
[arXiv:hep-th/9810250].

\bibitem{GS}
K. ~Galicki and S. Salamon, 
Geom. Ded. {\bf 63} (1996), 45-68.


\bibitem{Gray}
A. Gray,
T$\hat{\mbox{o}}$hoku Math. J. {\bf 21} (1969) 190-194,
J. Diff. Geom. {\bf 4} (1970), 283-309.


\bibitem{FKMS}
T. Friedrich, I. Kath, A. Moroianu, and U. Semmelmann,
J. Geom. Phys. {\bf 23} (1997), 259-286.



\bibitem{FK}
T. Friedrich and I. Kath,
J. Diff. Geom. {\bf 29} (1989), 263-279,
Comm. Math. Phys. {\bf 133} (1990), 543-561;
C. B\"{a}r,
Comm. Math. Phys. {\bf 154} (1993), 509-521.



\bibitem{Witten-G}
E.~Witten,
Nucl.\ Phys.\ B {\bf 186}, 412 (1981).



\bibitem{DFN}
R.~D'Auria, P.~Fre and P.~van Nieuwenhuizen,
Phys.\ Lett.\ B {\bf 136}, 347 (1984).



\bibitem{CR}
L.~Castellani and L.~J.~Romans,
Nucl.\ Phys.\ B {\bf 238}, 683 (1984).



\bibitem{PP}
D.~N.~Page and C.~N.~Pope,
Phys.\ Lett.\ B {\bf 147}, 55 (1984).



\bibitem{CDF}
L.~Castellani, R.~D'Auria and P.~Fre,
Nucl.\ Phys.\ B {\bf 239}, 610 (1984).



\bibitem{CRW}
L.~Castellani, L.~J.~Romans and N.~P.~Warner,
Nucl.\ Phys.\ B {\bf 241}, 429 (1984).


\bibitem{Romans}
L.~J.~Romans,
Phys.\ Lett.\ B {\bf 153}, 392 (1985).


\bibitem{DNP}
M.~J.~Duff, B.~E.~Nilsson and C.~N.~Pope,
Phys.\ Rept.\  {\bf 130}, 1 (1986).


\bibitem{Aloff}
S. Aloff and N. R. Wallach, 
Bull. Am. Math. Soc. {\bf 81} (1975), 93-97.





\bibitem{GV}
D.~Ghoshal and C.~Vafa,
Nucl.\ Phys.\ B {\bf 453}, 121 (1995)
[arXiv:hep-th/9506122].


\bibitem{OV}
H.~Ooguri and C.~Vafa,
Nucl.\ Phys.\ B {\bf 463}, 55 (1996)
[arXiv:hep-th/9511164].


\bibitem{ABKS}
O.~Aharony, M.~Berkooz, D.~Kutasov and N.~Seiberg,
JHEP {\bf 9810}, 004 (1998)
[arXiv:hep-th/9808149].


\bibitem{GKP}
A.~Giveon, D.~Kutasov and O.~Pelc,
JHEP {\bf 9910}, 035 (1999)
[arXiv:hep-th/9907178].


\bibitem{GK}
A.~Giveon and D.~Kutasov,
JHEP {\bf 9910}, 034 (1999)
[arXiv:hep-th/9909110];
A.~Giveon and D.~Kutasov,
JHEP {\bf 0001}, 023 (2000)
[arXiv:hep-th/9911039].



\bibitem{Pelc}
O.~Pelc,
JHEP {\bf 0003}, 012 (2000)
[arXiv:hep-th/0001054].


\bibitem{KutS}
D.~Kutasov and N.~Seiberg,
Phys.\ Lett.\ B {\bf 251}, 67 (1990);
D.~Kutasov,
{\it ``Some properties of (non)critical strings,''}
Lectures given at ICTP Spring School on String Theory and Quantum Gravity, Trieste, Italy, Apr 15-23, 1991,
arXiv:hep-th/9110041.


\bibitem{HK}
K.~Hori and A.~Kapustin,
JHEP {\bf 0211}, 038 (2002)
[arXiv:hep-th/0203147].



\bibitem{ES1}
T.~Eguchi and Y.~Sugawara,
Nucl.\ Phys.\ B {\bf 577}, 3 (2000)
[arXiv:hep-th/0002100].


\bibitem{Mizoguchi}
S.~Mizoguchi,
JHEP {\bf 0004}, 014 (2000)
[arXiv:hep-th/0003053];



\bibitem{Yamaguchi}
S.~Yamaguchi,
Nucl.\ Phys.\ B {\bf 594}, 190 (2001)
[arXiv:hep-th/0007069];
Phys.\ Lett.\ B {\bf 509}, 346 (2001)
[arXiv:hep-th/0102176];
JHEP {\bf 0201}, 023 (2002)
[arXiv:hep-th/0112004];


\bibitem{NN}
M.~Naka and M.~Nozaki,
Nucl.\ Phys.\ B {\bf 599}, 334 (2001)
[arXiv:hep-th/0010002].


\bibitem{Gepner}
D.~Gepner,
Phys.\ Lett.\ B {\bf 199}, 380 (1987);
Nucl.\ Phys.\ B {\bf 296}, 757 (1988).



\bibitem{GVW}
S.~Gukov, C.~Vafa and E.~Witten,
Nucl.\ Phys.\ B {\bf 584}, 69 (2000)
[Erratum-ibid.\ B {\bf 608}, 477 (2001)]
[arXiv:hep-th/9906070].

\bibitem{ShV}
A.~D.~Shapere and C.~Vafa,
{\it ``BPS structure of Argyres-Douglas superconformal theories,''}
arXiv:hep-th/9910182.


\bibitem{GGW}
S.~J.~Gates, S.~Gukov and E.~Witten,
Nucl.\ Phys.\ B {\bf 584}, 109 (2000)
[arXiv:hep-th/0005120].


\bibitem{EWY}
T.~Eguchi, N.~P.~Warner and S.~K.~Yang,
Nucl.\ Phys.\ B {\bf 607}, 3 (2001)
[arXiv:hep-th/0105194].



\bibitem{hybrid}
N.~Berkovits,
Nucl.\ Phys.\ B {\bf 431}, 258 (1994)
[arXiv:hep-th/9404162];
N.~Berkovits and C.~Vafa,
Nucl.\ Phys.\ B {\bf 433}, 123 (1995)
[arXiv:hep-th/9407190];
N.~Berkovits, C.~Vafa and E.~Witten,
JHEP {\bf 9903}, 018 (1999)
[arXiv:hep-th/9902098];
N.~Berkovits,
Nucl.\ Phys.\ B {\bf 565}, 333 (2000)
[arXiv:hep-th/9908041];
N.~Berkovits, S.~Gukov and B.~C.~Vallilo,
Nucl.\ Phys.\ B {\bf 614}, 195 (2001)
[arXiv:hep-th/0107140];


\bibitem{IK}
K.~Ito and H.~Kunitomo,
Phys.\ Lett.\ B {\bf 536}, 327 (2002)
[arXiv:hep-th/0204009].





\bibitem{ES3}
T.~Eguchi and Y.~Sugawara,
Phys.\ Lett.\ B {\bf 519}, 149 (2001)
[arXiv:hep-th/0108091].

\bibitem{SY2}
K.~Sugiyama and S.~Yamaguchi,
Phys.\ Lett.\ B {\bf 538}, 173 (2002)
[arXiv:hep-th/0204213].



\bibitem{SV}
S.~L.~Shatashvili and C.~Vafa,
Selecta \ Math. \ {\bf A1} 347 (1995)  
[arXiv:hep-th/9407025].


\bibitem{F-O}
J.~M.~Figueroa-O'Farrill,
Phys.\ Lett.\ B {\bf 392}, 77 (1997)
[arXiv:hep-th/9609113].



\bibitem{GN}
D.~Gepner and B.~Noyvert,
Nucl.\ Phys.\ B {\bf 610}, 545 (2001)
[arXiv:hep-th/0101116].



\bibitem{Noyvert}
B.~Noyvert,
JHEP {\bf 0203}, 030 (2002)
[arXiv:hep-th/0201198].



\bibitem{SY1}
K.~Sugiyama and S.~Yamaguchi,
Nucl.\ Phys.\ B {\bf 622}, 3 (2002)
[arXiv:hep-th/0108219].



\bibitem{JoyceCFT}
R.~Blumenhagen and V.~Braun,
JHEP {\bf 0112}, 006 (2001)
[arXiv:hep-th/0110232];
R.~Roiban and J.~Walcher,
JHEP {\bf 0112}, 008 (2001)
[arXiv:hep-th/0110302],
T.~Eguchi and Y.~Sugawara,
Nucl.\ Phys.\ B {\bf 630}, 132 (2002)
[arXiv:hep-th/0111012];
R.~Blumenhagen and V.~Braun,
JHEP {\bf 0112}, 013 (2001)
[arXiv:hep-th/0111048];
R.~Roiban, C.~Romelsberger and J.~Walcher,
{\it ``Discrete torsion in singular G(2)-manifolds and real LG,''}
arXiv:hep-th/0203272.


\bibitem{Joyce}
D. Joyce, 
{\it ``Compact Manifolds With Special Holonomy''} 
(Oxford University Press, 2000).



\bibitem{Seiberg-L}
N.~Seiberg,
Prog.\ Theor.\ Phys.\ Suppl.\  {\bf 102}, 319 (1990).
D.~Kutasov and N.~Seiberg,
Nucl.\ Phys.\ B {\bf 358}, 600 (1991).






\bibitem{Sotkov}
C.~Crnkovic, R.~Paunov, G.~M.~Sotkov and M.~Stanishkov,
Nucl.\ Phys.\ B {\bf 336}, 637 (1990).


\bibitem{KS}
Y.~Kazama and H.~Suzuki,
Nucl.\ Phys.\ B {\bf 321}, 232 (1989).


\bibitem{Sevrin}
A.~Van Proeyen,
Class.\ Quant.\ Grav.\  {\bf 6}, 1501 (1989);
A.~Sevrin and G.~Theodoridis,
Nucl.\ Phys.\ B {\bf 332}, 380 (1990).

\bibitem{Wolf}
J. A. Wolf, J. Math. Mech. {\bf 14} (1965), 1033.


\bibitem{ET}
T.~Eguchi and A.~Taormina,
Phys.\ Lett.\ B {\bf 200}, 315 (1988);
T.~Eguchi and A.~Taormina,
Phys.\ Lett.\ B {\bf 210}, 125 (1988).


\bibitem{odake}
S.~Odake,
Mod.\ Phys.\ Lett.\ A {\bf 4}, 557 (1989);
Int.\ J.\ Mod.\ Phys.\ A {\bf 5}, 897 (1990).


\bibitem{EOTY}
T.~Eguchi, H.~Ooguri, A.~Taormina and S.~K.~Yang,
Nucl.\ Phys.\ B {\bf 315}, 193 (1989).


\bibitem{HS}
Y.~Hikida and Y.~Sugawara,
JHEP {\bf 0210}, 067 (2002)
[arXiv:hep-th/0207124].





\bibitem{GRBL}
A.~Giveon and M.~Rocek,
JHEP {\bf 9904}, 019 (1999)
[arXiv:hep-th/9904024];
D.~Berenstein and R.~G.~Leigh,
Phys.\ Lett.\ B {\bf 458}, 297 (1999)
[arXiv:hep-th/9904040].

\bibitem{GKS}
A.~Giveon, D.~Kutasov and N.~Seiberg,
Adv.\ Theor.\ Math.\ Phys.\  {\bf 2}, 733 (1998)
[arXiv:hep-th/9806194].


\bibitem{SS}
A.~Schwimmer and N.~Seiberg,
Phys.\ Lett.\ B {\bf 184}, 191 (1987).


\bibitem{AGS}
R.~Argurio, A.~Giveon and A.~Shomer,
JHEP {\bf 0004}, 010 (2000)
[arXiv:hep-th/0002104];
JHEP {\bf 0012}, 025 (2000)
[arXiv:hep-th/0011046].



\bibitem{CHS}
C.~G.~Callan, J.~A.~Harvey and A.~Strominger,
Nucl.\ Phys.\ B {\bf 359}, 611 (1991),
Nucl.\ Phys.\ B {\bf 367}, 60 (1991).


\bibitem{GHM}
R.~Gregory, J.~A.~Harvey and G.~W.~Moore,
Adv.\ Theor.\ Math.\ Phys.\  {\bf 1}, 283 (1997)
[arXiv:hep-th/9708086].






\bibitem{CFTtext}
P. Di Francesco, P. Mathieu and D. S\'{e}n\'{e}chal,
{\it ``Conformal Field Theory,''}
Graduate Texts in Contemporary Physics, Springer.


\bibitem{CIZ}
A.~Cappelli, C.~Itzykson and J.~B.~Zuber,
Nucl.\ Phys.\ B {\bf 280}, 445 (1987),
Commun.\ Math.\ Phys.\  {\bf 113}, 1 (1987);
A.~Kato,
Mod.\ Phys.\ Lett.\ A {\bf 2}, 585 (1987).



\bibitem{ES2}
T.~Eguchi and Y.~Sugawara,
Nucl.\ Phys.\ B {\bf 598}, 467 (2001)
[arXiv:hep-th/0011148].



\bibitem{PZT}
L.~A.~Pando Zayas and A.~A.~Tseytlin,
Class.\ Quant.\ Grav.\  {\bf 17}, 5125 (2000)
[arXiv:hep-th/0007086].


\bibitem{GMM}
E.~Guadagnini, M.~Martellini and M.~Mintchev,
Phys.\ Lett.\ B {\bf 194}, 69 (1987);
E.~Guadagnini,
Nucl.\ Phys.\ B {\bf 290}, 417 (1987).


\bibitem{BS}
R. Bryant and S. Salamon, 
Duke Math. J. {\bf 58} (1989) 829.


\bibitem{GPP}
G.~W.~Gibbons, D.~N.~Page and C.~N.~Pope,
Commun.\ Math.\ Phys.\  {\bf 127}, 529 (1990).


\bibitem{LVW}
W.~Lerche, C.~Vafa and N.~P.~Warner,
Nucl.\ Phys.\ B {\bf 324}, 427 (1989).



\bibitem{EFGT}
S.~Elitzur, O.~Feinerman, A.~Giveon and D.~Tsabar,
Phys.\ Lett.\ B {\bf 449}, 180 (1999)
[arXiv:hep-th/9811245].


\bibitem{2dgrav}
V.~G.~Knizhnik, A.~M.~Polyakov and A.~B.~Zamolodchikov,
Mod.\ Phys.\ Lett.\ A {\bf 3}, 819 (1988);
F.~David,
Mod.\ Phys.\ Lett.\ A {\bf 3}, 1651 (1988);
J.~Distler and H.~Kawai,
Nucl.\ Phys.\ B {\bf 321}, 509 (1989).




\bibitem{2DBH}
E.~Witten,
Phys.\ Rev.\ D {\bf 44}, 314 (1991);
G.~Mandal, A.~M.~Sengupta and S.~R.~Wadia,
Mod.\ Phys.\ Lett.\ A {\bf 6}, 1685 (1991);
I.~Bars and D.~Nemeschansky,
Nucl.\ Phys.\ B {\bf 348}, 89 (1991);
S.~Elitzur, A.~Forge and E.~Rabinovici,
Nucl.\ Phys.\ B {\bf 359}, 581 (1991).




\bibitem{DM}
K.~Dasgupta and S.~Mukhi,
Nucl.\ Phys.\ B {\bf 551}, 204 (1999)
[arXiv:hep-th/9811139],
JHEP {\bf 9907}, 008 (1999)
[arXiv:hep-th/9904131].

\bibitem{GT}
S.~Gukov and D.~Tong,
Phys.\ Rev.\ D {\bf 66}, 087901 (2002)
[arXiv:hep-th/0202125];
JHEP {\bf 0204}, 050 (2002)
[arXiv:hep-th/0202126].


















\end{thebibliography}
\end{document}